\documentclass[aps,pre,twocolumn,float]{revtex4-1}
\usepackage{amsmath,bm,epsfig}

\newcommand{\B}[1]{{\bm{#1}}}%% Bold Roman & Greek Lower & Upper Case
\newcommand{\C}[1]{{\mathcal{#1}}}    %%   Calligrapfic Upper case
\usepackage[dvips]{color}
\begin{document}
%%%%%%%%%%%%%%%%%%%%%%

\title{The sandpile revisited: computer assisted determination \\of constitutive 
relations and the breaking of scaling}
\author{H. George E. Hentschel$^*$, Prabhat K. Jaiswal, Chandana Mondal, Itamar 
Procaccia and Jacques Zylberg}
\affiliation{Department of Chemical Physics, The Weizmann Institute of Science,  
Rehovot 76100, ISRAEL\\
$^*$Department of Physics, Emory University, Atlanta GA 30322, USA}
\begin{abstract}
We revisit the problem of the stress distribution in a frictional sandpile under 
gravity, equipped with
a new numerical model of granular assemblies with both normal and tangential 
(frictional) inter-granular forces.
Numerical simulations allow a determination of the spatial dependence of all the 
components of the stress
field as a function of systems size, the coefficient of static friction and the 
frictional interaction
with the bottom surface. Our study clearly demonstrates that interaction with 
the bottom surface plays a crucial role in the formation of a pressure dip under 
the apex of a granular pile. Basic to the theory of sandpiles
are assumptions about the form of scaling solutions and constitutive relations 
for cohesive-less hard grains for which no typical scale is available. We find 
that these constitutive relations must be modified; moreover for smaller 
friction coefficients and smaller piles these scaling
assumptions break down in the bulk of the sandpile due to the presence of length 
scales that must be carefully identified. After identifying the crucial
scale we provide a predictive theory to when scaling solutions are expected to 
break down. At the bottom of the pile the scaling assumption always breaks, due 
to the different interactions with the bottom surface. The consequences for 
measurable quantities like the pressure distribution and shear stress at the 
bottom of the pile are discussed. For example
one can have a transition from no dip in the base-pressure to a dip at the 
center of the pile as a function
of the system size.
\end{abstract}
\maketitle

\section{Introduction}

Piles of granular matter are all around us, from apples in the market stalls to 
dunes near the beach. The ubiquity of such aggregates that are typified by 
sandpiles have raised the interest of scientists for a long time, driven 
theoretically to understand the shape of such piles (angle of repose) and the 
distribution of normal and shear stresses. Needless to say these issues are also 
technologically of high interest from the engineering of silos to the 
wind-induced migration of sand dunes. Indeed, experimental knowledge has 
accumulated for over a century \cite{20HF,38Hou,56Tro,81SN,99VHCB,13AOCRZ}. 
Among the many findings relating to the angle of repose
and the stability of such piles, experiments indicated the non-intuitive result 
that the pressure at the bottom of the pile
does not necessarily maximize at the center. Rather, there may appear  a dip in 
the pressure at the center with a ring of
maxima (for piles in 3 dimensions) some radius away from the center point. In 
2-dimensional piles there may be two maxima
in the pressure some distance away from the dip in the center. The existence of 
such a dip may depend on material parameters and growth protocols like the 
piling history\cite{14ZY}, the shape of the grains \cite{93PB}, and as we show 
below, also on the size of the pile and coefficient of static friction $\mu$. 
Computational investigations corroborated these experimental results 
\cite{92LCH,94BP}.
The explanation of the dip in the pressure is still not fully settled; in a 
recent publication \cite{14ZY} it was even questioned whether classical 
continuum theories can predict such dips at all. Below we will show that 
computer assisted continuum theories can indeed result
in a dip in the pressure under the apex of the pile.

Modern theoretical developments began with Edwards' and Oakeshott's proposition 
of the concept of arching \cite{89EO}. This concept underlined the relevance of 
the directional properties of the field of principal axes of the stress tensor. 
As pointed out in later contributions \cite{95BCC,96WCCB,97WCC} this concept 
requires an understanding of the spatial solution of
the stress tensor, a problem whose mechanics is under-determined. Accordingly, a 
constitutive relation has to be selected. Since this relation is not dictated by 
equilibrium mechanics, this selection may be dangerous as it may depend on 
details of the friction mechanism, the shape of the particles, the protocol of 
growing the sandpile \cite{99VHCBC} etc.

The advent of supercomputers allows us to revisit these interesting issues 
equipped with a granular model
where the effects of friction coefficient, the protocol of growth and the system 
size can be studied carefully, and where the stress field can be measured 
accurately. As shown below we find that generically the constitutive relations 
are hard to guess a-priori, and moreover, they may change with the growth of the 
pile. For example we will show that a dip in the
pressure at the bottom surface under the apex of the pile may develop with the 
growth of the pile \cite{01BCLO}.

A major theoretical assumption made in analyzing the equations of static 
equilibrium \cite{95BCC,96WCCB,97WCC} is the existence of scaling solutions
which for a two-dimensional sandpile in $(x,z)$ coordinates depend only on a 
scaling variable $S\equiv x\tan \phi/z$ where
$\phi$ is the angle of repose. The coordinates $x$ and $z$ are measured from the 
apex of the pile with $z$ in the direction of gravitational force. The scaling 
solutions are only tenable when there is no typical length scale
in the sandpile other than the system size. We show below that this assumption 
is broken for small friction coefficients and for small number of grains. One of 
our tasks is to identify the most relevant length-scale and correlate it with 
the breaking
of scaling and its inevitable consequences. More importantly, even when there 
exist scaling solutions they are not necessarily of the type
assumed in the literature as is shown below. Interestingly, often scaling is 
only weakly broken in the bulk of the pile, but strongly broken at the bottom 
where interactions with the supporting surface may differ from the bulk 
interactions. We will show
that the appearance of a pressure dip in the center may very well be related to 
this strong breaking of scaling.

The structure of the paper is as follows: In Sect.~\ref{model} we describe our 
numerical model and the results of
numerical simulations including the dependence of the angle of repose on the 
static friction coefficients and the profiles
of pressure and shear stresses everywhere in the pile. In 
Sect.~\ref{constitutive} we briefly review the theoretical background available 
in the literature. In Sect~\ref{theory} we explain the theoretical modification 
necessary to get a closer agreement with the simulation results and to provide 
accurate estimates of the stress and pressure profiles that are of interest. 
Sect.~\ref{scaling} examines the observed breakdown in scaling and the necessity 
for incorporating new length-scales to account for the observed phenomena. 
Sect.~\ref{summary} offers a summary, conclusions, and some remarks on the road 
ahead.

\section{The Model and Simulation Results}
\label{model}
\subsection{The model}
Many models of frictional amorphous materials use time dependent dynamics. In 
our work we opt to derive a model
that can provide complete mechanical equilibration such that at rest the force 
and the torque on each grain vanishes.
To implement an amorphous (non-crystalline) two-dimensional sandpile we 
construct a model consisting of a 50:50 binary system of
disks of two sizes, one with a smaller radius $\sigma_s=\lambda$ and the other 
with a larger radius $\sigma_\ell=1.4\lambda$. At every contact between any two 
disks $i$ and $j$ we assign a normal and a tangential
linear spring aligned in parallel and perpendicularly to the radius vector 
connecting their center of masses. The normal and tangential spring constants 
are $K_N$ and $K_T$, respectively. The position of the center of mass of the 
$i$th disk is denoted as $\B R_i$ while the vector distance\ between the center 
of mass of two particles in contact is denoted as $\B r^{ij}$. The unit vector 
in the direction of $\B r^{ij}$ is $\hat {\B r}^{ij}$.

The force law for the normal spring, $\B f^{ij}_N=K_N \B n^{ij}$, responds to 
the amount of overlap between particles $i$ and $j$, $\B 
n^{ij}=(\sigma_i+\sigma_j)\hat {\B r}^{ij}-\B r^{ij}$, as determined by the 
system configuration. The tangential spring has an analogous force law, $\B 
f_T^{ij}=K_T \B t^{ij}$, however the value of $t^{ij}$ is not
readily available from the system configuration. The tangential spring 
elongation $t^{ij}$ is dependent on the system's history and needs thus to be 
tracked along the system's path. Consider particles $i$ and $j$; when this pair 
of particles is coming into contact the tangential spring is created and by 
definition is at its rest length, $t^{ij}=0$. When grazing, the spring is loaded 
with respect to the
degree of grazing, $\Delta 
t^{ij}=\sigma_i\Delta\Phi^{ij}+\sigma_j\Delta\Phi^{ji}$. Here $\Delta\Phi^{ij}$ 
denotes change in angle of the contact point of disc $i$ on disc $j$. For 
computing the degree of relative rotation we consider the radius to be invariant 
as long as the compression is small. The expression $\sigma_i\Delta\Phi^{ij}$ 
therefore captures the arc-length and direction along which the point of contact 
with disc $j$ has moved about particle $i$. During the simulation the actual 
value of $t^{ij}$ is obtained by summing up the incremental
values $\Delta t^{ij}$ until the Coulomb limit is reached.
In order to fully define the behavior of particle interaction at the contact it 
is necessary to differentiate between two types of tangential motions which are 
{\it rolling} and {\it grazing}. Two particles rolling on top of each other do 
not load the spring, $t^{ij}$ remains unchanged. Both types of interaction are 
taken into account in the present model. Finally, particles $i$ and $j$
share the same springs and it therefore follows, by Newton's third law, that  
$\B t^{ij}=-\B t^{ji}$ and $\B n^{ij}=-\B n^{ji}$.

The frictional slips are dominated by Coulomb's law; the maximal deviation of 
the tangential spring from its rest length is defined by $f_{T,{\rm 
max}}^{ij}=\mu f_N^{ij}$, where $\mu$ is the friction coefficient. When 
$f_T^{ij}=f_{T,{\rm max}}^{ij}$, then the contact
breaks and a slip event occurs, redefining the anchoring point of the tangential 
spring on both particles $i$ and $j$. Following experimental evidence 
\cite{Fineberg} we model the slip event by giving up a fixed portion, chosen 
below as 20\%, of the tangential
loading. In other words, upon reaching the Coulomb limit we set
$t_{\rm new}^{ij}=0.8 t_{\rm old}^{ij}$ (representing
the movement of the tangential spring's anchor on the surface of both particles 
$i$ and $j$).

To construct a sandpile we start with a horizontal ``floor" at $h=0$ along the 
$x$-direction, with one small disk positioned
at $x=0$. We then add alternating small and large disks always at $x=0$ with 
some noise $\delta x$ chosen with a uniform distribution in the range 
$[-0.1,0.1]$. The particle is positioned at a
small clearance in its height that guarantees zero overlap with the highest disk 
in the pile.  After each addition of a disk,
we perform conjugate gradient minimization to reach mechanical equilibrium where 
the forces and torques on each particle sum up to zero. The interaction of the 
particles with the
floor is at our disposal, and we chose to have the same normal law but with with 
a higher value of $K_T$. There is no reason to assume that the grains interact 
with the floor as they do among themselves. Most of the
simulations described below are done with a value of $K_T$ for the floor 
interaction that is double the value in the bulk. We do however test different 
ratios of the tangential force constants, and see below for details. Finally,
the acceleration due to gravity was chosen in the following way: to get a 
realistic sandpile we should choose the gravitational energy $mg\lambda$ of one 
particle at height $\lambda$ to be of the order of the elastic energy 
$\frac{1}{2}K_N (\delta\lambda)^2$ where $\delta$ is the amount of compression. 
For $m=1$ for the small particle (and $m=1.96$ for the large particle), 
$K_N=0.1$ and $K_T=0.05$ we expect $\delta$ not to exceed, say $10^{-1}$, and 
thus $g$ should be
chosen of the order of $g=0.001$.  With this value of $g$ and all the other 
choices
of parameters we get sandpiles that appear physical.
%%%%%%%%%%%%%%%%%%%%%%%%%%%%%%%%%%%%%%%%%%%
\begin{figure}
\includegraphics[width=0.5\textwidth]{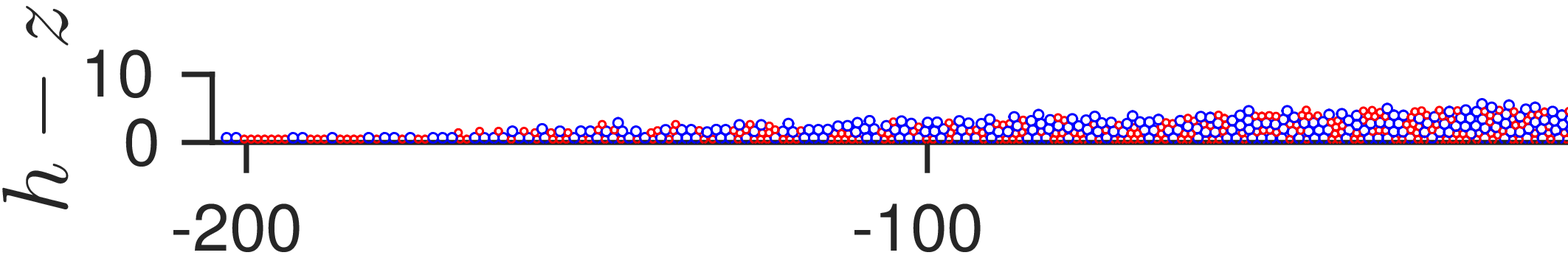}
\includegraphics[width=0.5\textwidth]{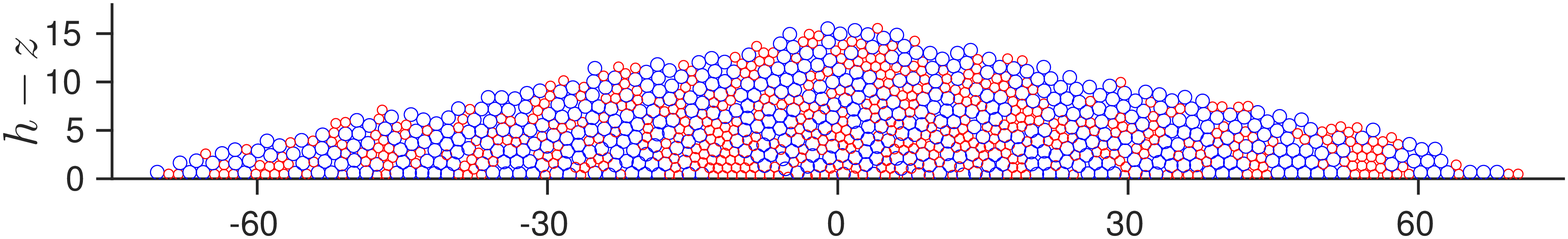}
\includegraphics[width=0.5\textwidth]{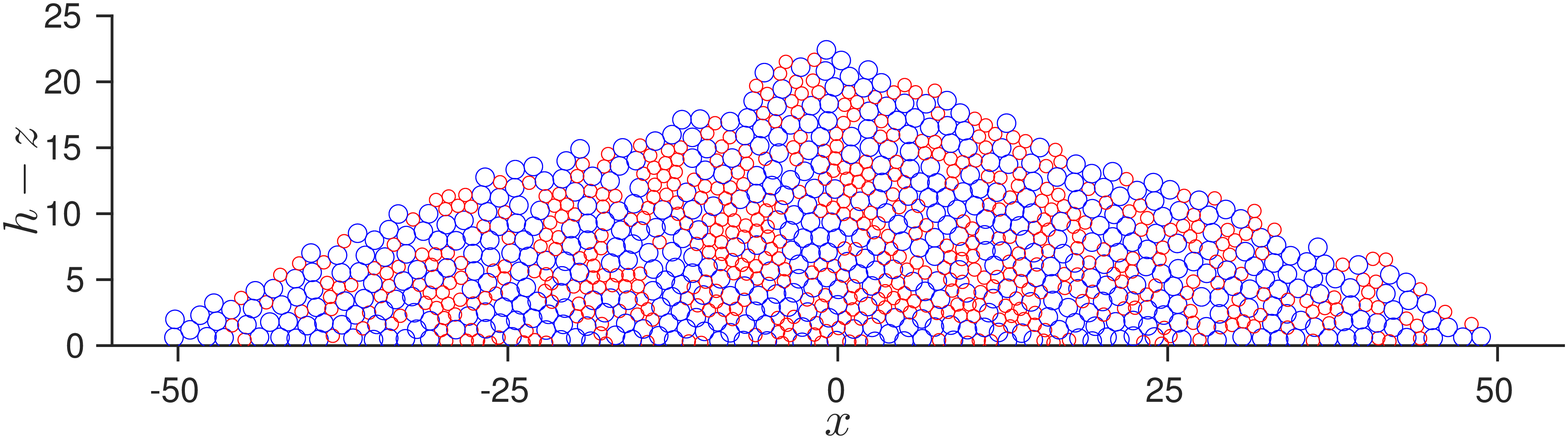}
\caption{Typical sandpiles grown by the model which employs circular ``grains" 
of two sizes. The friction coefficients are $\mu=0.01$ (upper panel),
$\mu=0.1$ (middle panel) and $\mu=1$ (lower panel).}
\label{sandpiles}
\end{figure}
%%%%%%%%%%%%%%%%%%%%%%%%%%%%%%%%%%%%%%%%%%%%%%%%%%%%%%%%%%%%%%%%%%%
\subsection{Results of simulations}

\begin{figure}
\includegraphics[width=0.35\textwidth]{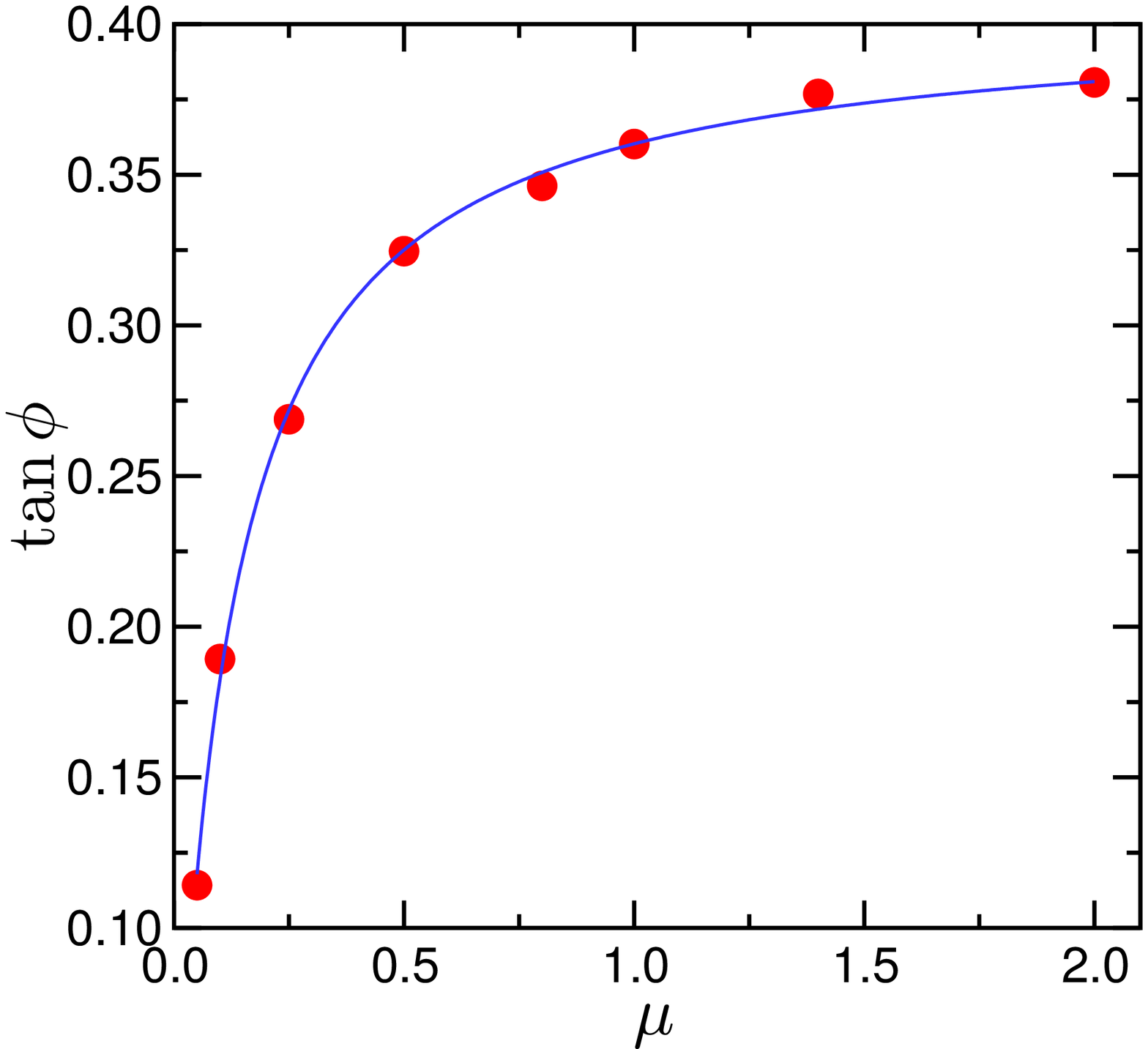}
\includegraphics[width=0.35\textwidth]{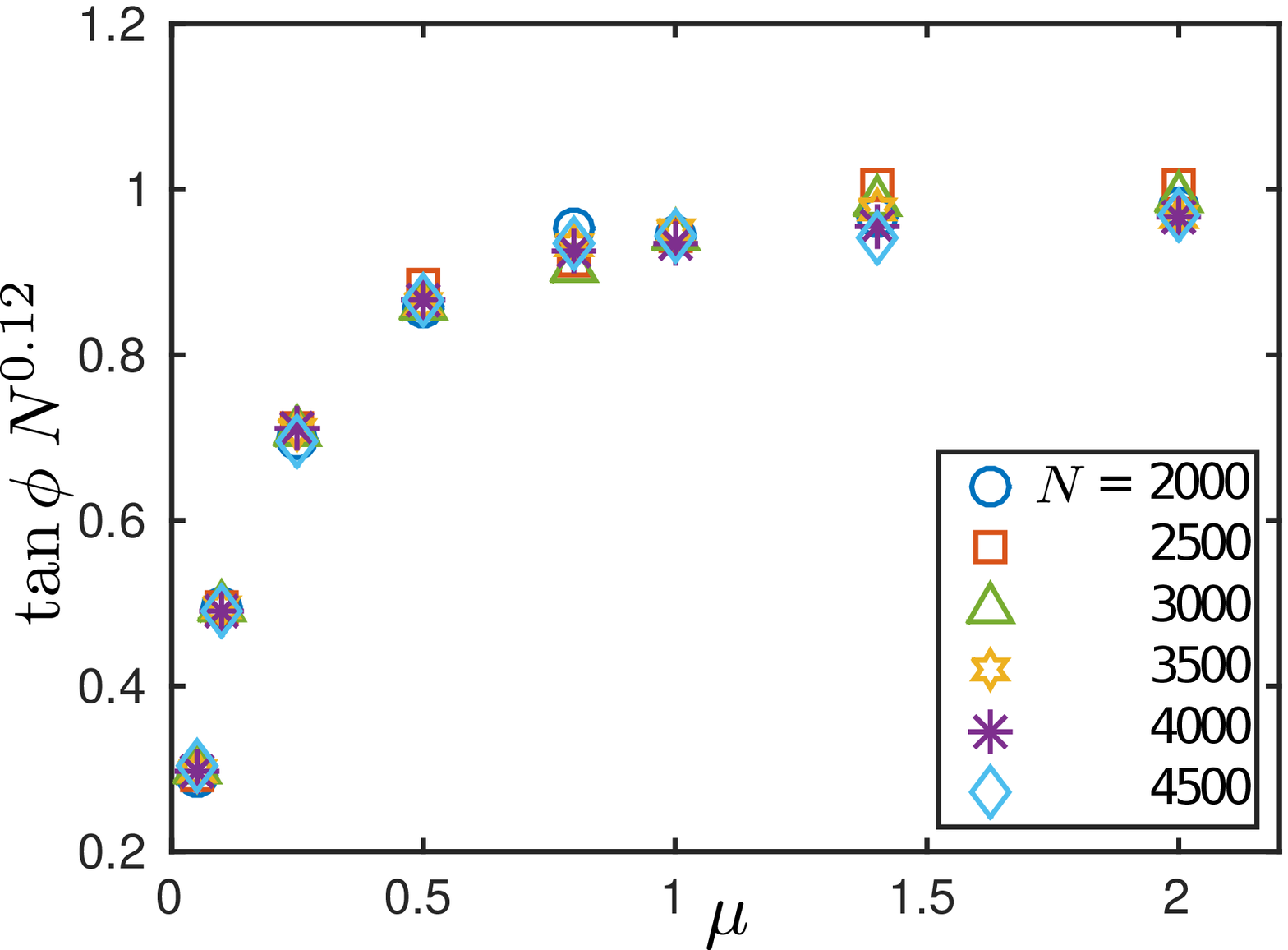}
\caption{Upper panel: $\tan \phi$ as a function of the friction coefficient 
$\mu$ for a given
system size $N=3000$. The fitting function
in continuous blue line is $\tan\phi \approx C_1 \mu/(1+C_2 \mu)$, cf. 
Eq.~\ref{pade}.
Lower panel: data collapse of $\tan \phi$ for different values of system size 
$N$ as a function
of $\mu$. It is possible that for much larger values of $N$ the angle or repose 
reaches an asymptotic
value.}
\label{tanphi}
\end{figure}

Typical results for sandpiles are shown in Fig.~\ref{sandpiles} for a number of 
values of the friction coefficient $\mu$. Each
of these realizations can be used to define an angle of repose $\tilde \phi$ by 
best fitting to a triangular shape, but of course this angle of repose is a 
random variable with fluctuations that depend on $N$. We determine the average 
angle of repose $\phi$
by averaging $\tilde \phi$ over ${\C N}$ realizations. For example we created 
sandpiles with $\C N=718, 1123, 555$ for $\mu=0.01, 0.1, 1$, respectively. The 
relation between the angle of repose $\phi$ at a given size $N$ and $\mu$ for 
given
material parameters $K_N, K_T$ and for the fixed circular disks in this model is 
shown in Fig.~\ref{tanphi}. One sees
that $\tan \phi$ starts as a linear function of $\mu$ for small $\mu$, $\tan\phi 
\approx C_1\mu$ where $C_1\approx 2.5$. At larger
values of $\mu$, $\tan\phi$ saturates with a maximal angle of repose $\phi_{\rm 
max}\approx 24^\circ$. The function shown
in Fig.~\ref{tanphi} can be fitted using a Pad\'e approximation as
\begin{equation}
\tan\phi \approx \frac{C_1 \mu}{1+C_2 \mu} \ , \quad \frac{C_1}{C_2} = \tan 
\phi_{\rm max} \ .
\label{pade}
\end{equation}
%%%%%%%%%%%%%%%%%%%%%%%%%%%%%%%%%%%%%%%%%
%%%%%%%%%%%%%%%%%%%%%%%%%%%%%%%%%%

It should be noted that for systems sizes simulated here the angle of repose is 
a weak function of the
system size $N$, decreasing approximately like $N^{-0.12}$ for all the values of 
$\mu$. In other words,
for different system sizes $N$ and different values of $\mu$ one can find a 
master curve for which the angle
of repose shows data collapse by plotting $\tan \theta \times N^{0.12}$ vs. 
$\mu$, see Fig.~\ref{tanphi} lower
panel. A similar weak dependence of the angle of repose on system size for small 
size sandpiles had been found
numerically before, cf.  Ref.~\cite{94BP}.
%%%%%%%%%%%%%%%%%%%%%%%%%%%%%%%%%%%%%%%%%%%%%%%%%%%%%%%%%%%%%%%%%%%
\begin{figure}
\includegraphics[width=0.5\textwidth]{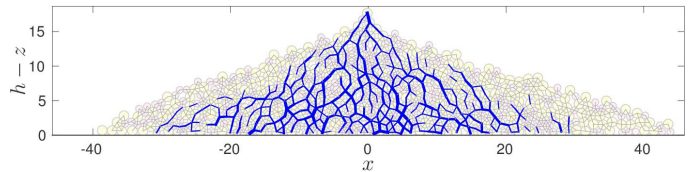}
\caption{Typical appearance of the force chains in a sandpile of size $N=600$ 
and $\mu=0.35$.}
\label{forcechains}
\end{figure}
%%%%%%%%%%%%%%%%%%%%%%%%%%%%%%%%%%%%%%%

An interesting and often studied characteristic of granular compressed media is 
the geometry of force chains.
To present these one needs to choose a threshold for the inter-particle forces 
$\B f_{ij}$ at the contacts and
plot only those forces that exceed the threshold. This is somewhat arbitrary and 
is in the eyes of the beholder.
Choosing the force chains to be system spanning \cite{Somfai} we get typical 
results as shown in Fig.~\ref{forcechains}.
This figure underlines the existence of a typical scale that will be discussed 
below, i.e. the typical distance
between bifurcations of the force chains.

%%%%%%%%%%%%%%%%%%%%%%%%%%%%%%%%%%%%%%%%%%%%%%%%%
\begin{figure}
\includegraphics[width=0.35\textwidth]{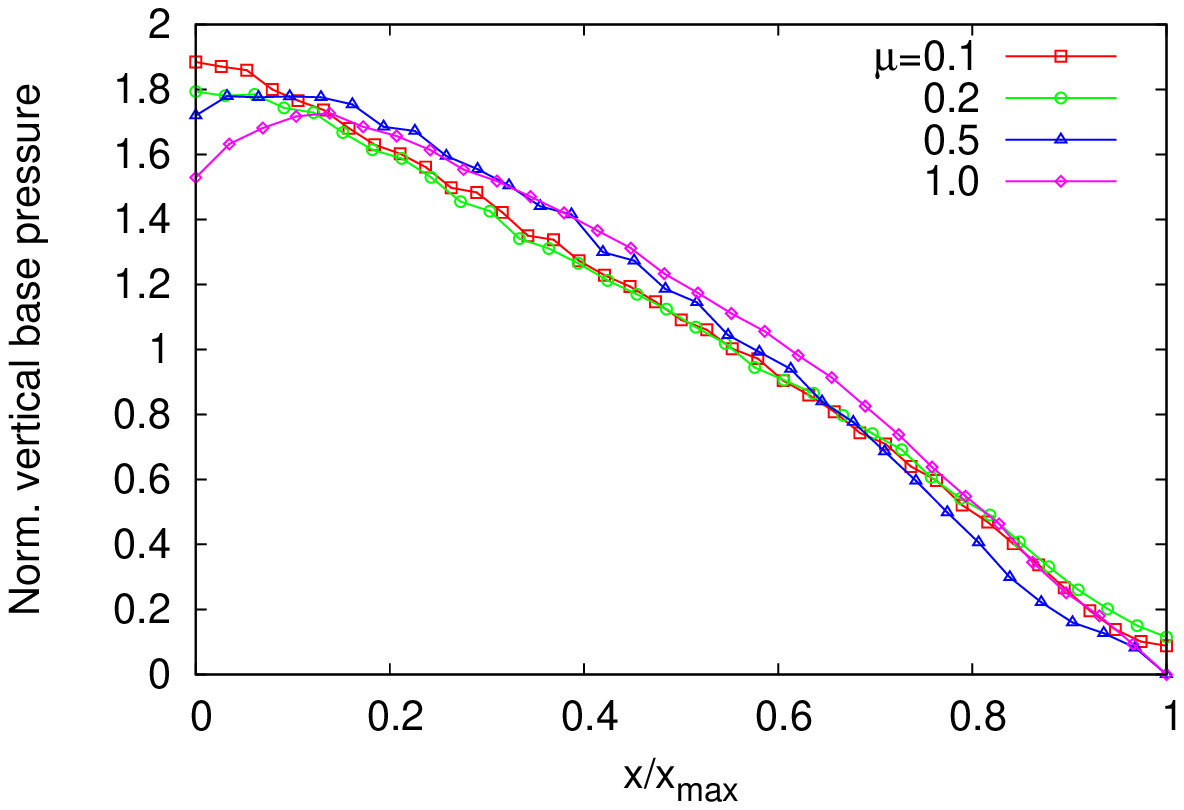}
\includegraphics[width=0.35\textwidth]{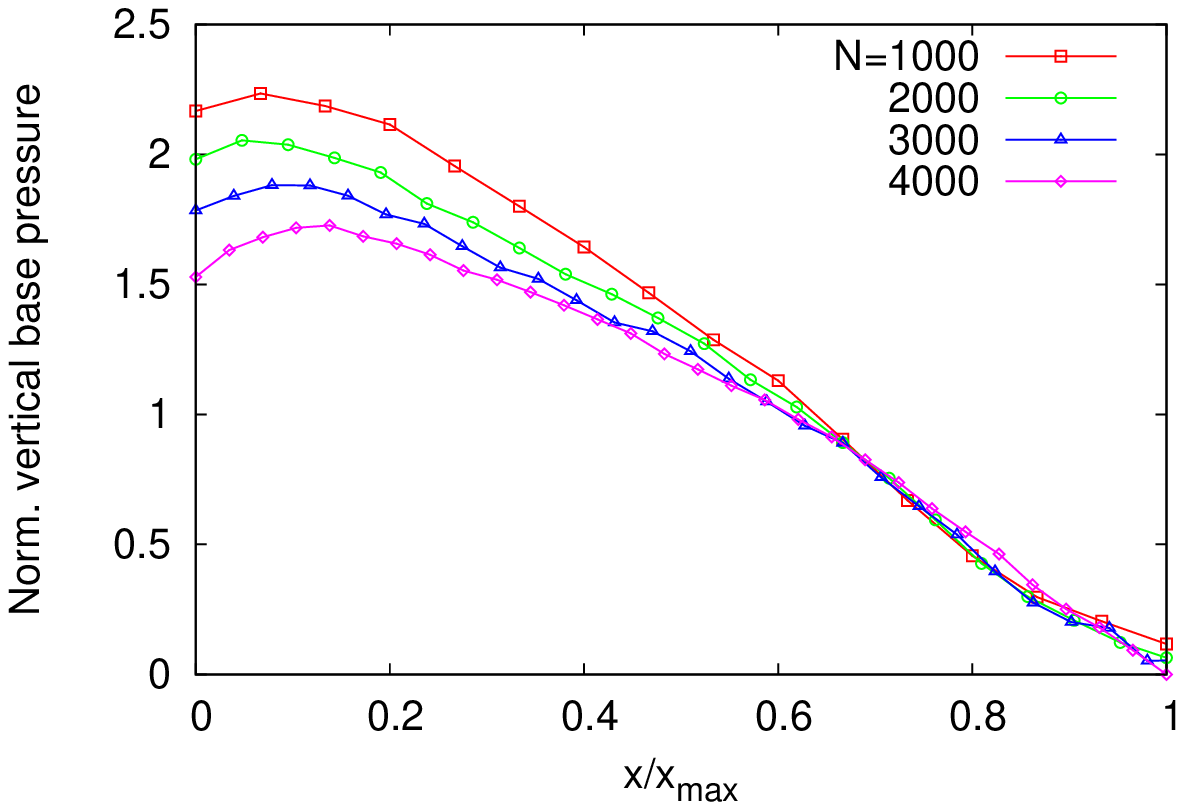}
\caption{Base pressure distribution normalized by $Mg/(2x_{max})$. Upper panel: 
systems consisting of $4000$ grains with different $\mu$. Lower panel: $\mu=1$ 
but with different number of grains, $N$.}
\label{basep}
\end{figure}
\begin{figure}
\vskip 0.3cm
\includegraphics[width=0.35\textwidth]{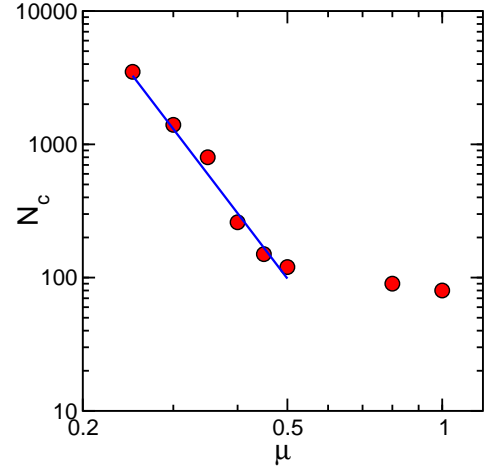}
\caption{Number of grains below which no pressure dip has been observed, as a 
function of $\mu$.}
\label{Nc}
\end{figure}

%%%%%%%%%%%%%%%%%%%%%%%%%%%%%%%%%%%%%%%%%%%%%%
Fig.~\ref{basep} shows the pressure distribution along the bottom (base 
pressure) of the pile. The base pressure is defined to be the amplitude of the 
force acting on each grain fixed on the bottom and averaged in symmetric cells 
of dimension $4.0\lambda$. The base pressure is normalized by $Mg/(2x_{max})$ 
where $M$ is the total mass of the pile and $x_{max}$ is the maximum horizontal 
spread from the center of the pile. In the upper panel of Fig.~\ref{basep} we 
show that there is a transition from no-dip to dip as $\mu$ is increased. The 
number of grains $N$ is fixed at $4000$ for all the curves. In the lower panel 
of Fig.~\ref{basep} we show that even for a fixed $\mu$ ($=1$ in the plot) the 
dip-height reduces very sharply with decreasing $N$. For $\mu=1$ we see pressure 
dip even for very small piles $N=500$. However, for lower $\mu$ we observe a 
no-dip to dip transition as $N$ is increased. In Fig.~\ref{Nc} we plot the 
critical number of grains $N_c$ below which no pressure dip is observed at the 
corresponding $\mu$.

%%%%%%%%%%%%%%%%%%%%%%%%%%%%%%%%%%%%%%%%%%%%%%%%%
%%%%%%%%%%%%%%%%%%%%%%%%%%%%%%%%%%%%%%%%%%%%%%

%%%%%%%%%%%%%%%%%%%%%%%%%%%%%%%%%%%%%%%%%%%%%%%%%
\begin{figure}
\includegraphics[width=0.35\textwidth]{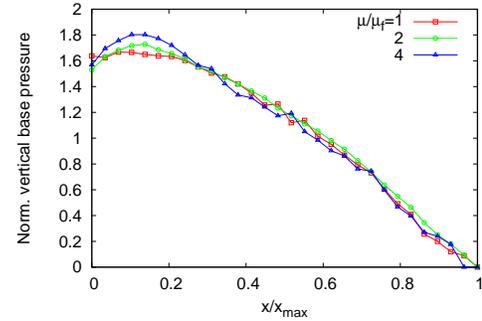}
\caption{Base pressure distribution normalized by $Mg/(2x_{max})$ for three 
different particle-floor friction coefficient ${\it viz}$ $\mu_f=1.0, 2.0, 4.0$. 
The particle-particle friction coefficient $\mu=1$.}
\label{muf}
\end{figure}
%%%%%%%%%%%%%%%%%%%%%%%%%%%%%%%%%%%%%%%%%%%%%%

Another striking observation is that the dip in the base-pressure depends 
strongly on the floor to particle friction coefficient. The dip height increases 
with the increase in the floor-particle friction coefficient which we denote by 
$\mu_f$. This is shown in Fig.~\ref{muf} where we plot the distribution of 
normalized base pressure for three different floor-particle friction coefficient 
${\it viz}$ $\mu_f=1.0, 2.0, 4.0$, particle-particle friction coefficient 
$\mu=1$ and the number of grains being $N=4000$ for all the cases. In all the 
simulations in which we do not report a different value we have used 
$\mu_f=2\mu$.

%%%%%%%%%%%%%%%%%%%%%%%%%%%%%%%%%%%%%%%%%%%%%%%%%
\begin{figure}
\includegraphics[width=0.5\textwidth]{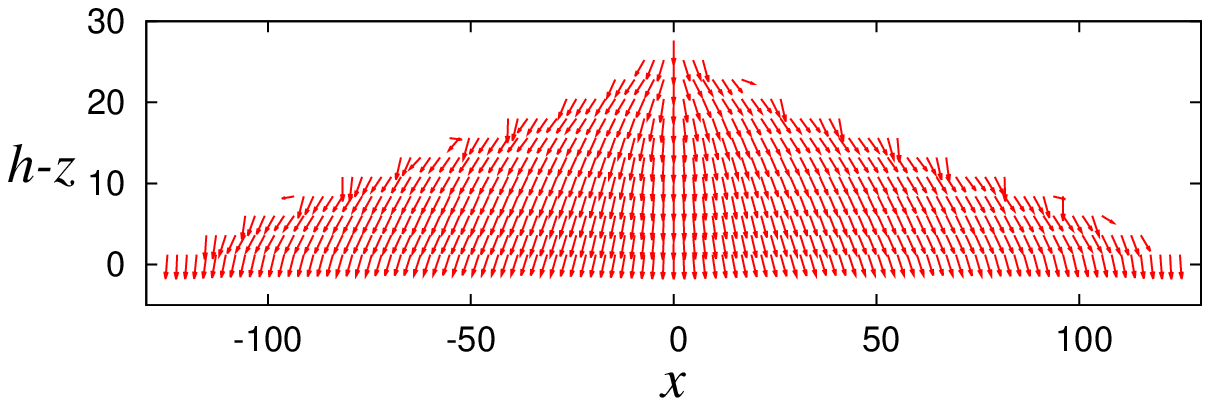}
\includegraphics[width=0.5\textwidth]{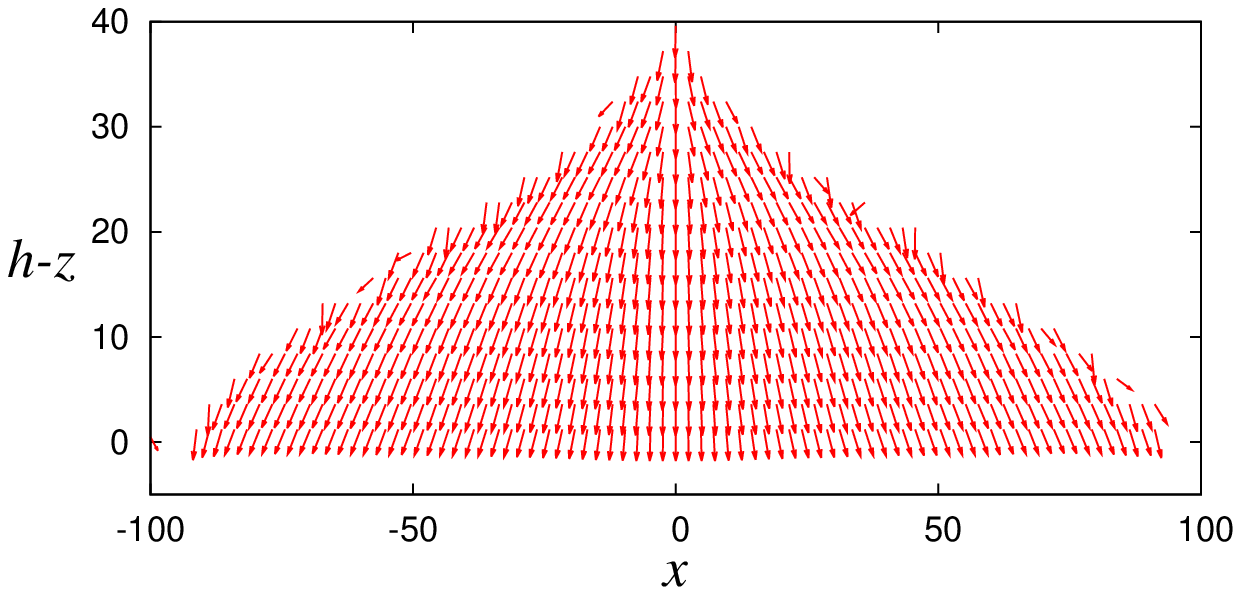}
\caption{Direction of the principal axis $\Psi(x,z)$ of the stress in the 
sandpile for two different
values of the friction coefficient. Upper panel: $\mu=0.1$ Lower panel $\mu=1$. 
A similar
figure for $\mu=0.01$ is not shown since it does not display well.}
\label{psidir}
\end{figure}
%%%%%%%%%%%%%%%%%%%%%%%%%%%%%%%%%%%%%%%%%%%%%%
A quantity of crucial theoretical interest which is used below in the context of 
constitutive relations is the direction of the principal axis of the stress 
tensor in the pile. We define $\Psi(x,z)$ as the angle between the principal 
axis and the $z$-axis.
To obtain
our numerical calculation of $\Psi$ we consider the aforementioned $\C N$ copies 
of sandpiles with a given value of $\mu$,
and system size $N$. For each realization we compute the stress field on 
plaquettes of sizes $2.4\lambda\times 2.4\lambda$ using
the standard algorithms \cite{07IMPS}. We average the obtained stress field in 
each plaquette over all the realizations. Finally, we determine the principal
axis $\Psi(x,z)$ using Eqs.~(\ref{defpsi}) below. Typical
results for $\mu=0.1$ and 1 are shown in Fig~\ref{psidir}. The image for 
$\mu=0.01$ does not display well.
The reader should note the very strong fluctuations, even after averaging on 
many independent samples of $\Psi(x,z)$
on the surface of the sand piles. We will argue below that these fluctuation are 
responsible in part for destroying the scaling
assumptions.

\section{Theoretical background}
\label{constitutive}

An extremely useful and important review is provided by Ref.~\cite{97WCC}. For 
the two-dimensional sandpile (which corresponds
to our simulations) the fundamental equations for the stress field at 
equilibrium read
\begin{eqnarray}
\partial_x \sigma_{xx} +\partial_z\sigma_{xz} &=&0 \nonumber\\
\partial_x \sigma_{xz} +\partial_z\sigma_{zz} &=&mg/\lambda^2 \ ,
\label{equil}
\end{eqnarray}
where $mg$ is the gravity force on a grain of mass $m$ and area $\lambda^2$.
The coordinate $z$ is measured from the apex of the pile and $x$ is measured 
from the symmetry axis. Note that
these equations are under-determined since there are two equations for the three 
independent components of the
stress tensor $\sigma_{xx}, \sigma_{xz}=\sigma_{zx}$ and $\sigma_{zz}$. Using 
the trace and determinant of the stress
tensor as invariants one can find the angle of inclination $\Psi$ between the 
principal axis of the stress and
the $z$ axis. Defining the invariants
\begin{eqnarray}
P&\equiv& \case{1}{2}(\sigma_{xx}+\sigma_{zz})\nonumber\\
R^2&\equiv& \case{1}{4}(\sigma_{zz}-\sigma_{xx})^2+\sigma^2_{zx} \ ,
\label{inv}
\end{eqnarray}
one can then write
\begin{eqnarray}
\sigma_{xx} &=& P-R\cos(2\Psi)\ , \nonumber\\
\sigma_{zz} &=& P+R\cos(2\Psi)\ , \nonumber\\
\sigma_{xz}&=& R\sin(2\Psi)\ .
\label{defpsi}
\end{eqnarray}
The maps of principal axes shown in Sect.~\ref{model} were computed with the 
help of these equations.

Discussing the stability of the sand pile, we recall that the simulations employ 
three independent material
parameters, i.e. $K_N$, $K_T$ and $\mu$. It is customary to introduce a 
phenomenological material constant,
sometime denoted as $\tan\phi^*$ which is referred to as the angle of internal 
friction and used in
the limit of stability
\begin{equation}
|\sigma_{nt}| \le \tan\phi^* \sigma_{nn} \ ,
\label{stability1}
\end{equation}
where $\sigma_{nn}$ and $\sigma_{nt}$ are the normal and tangential components 
of the stress on any chosen
plane. We find in our simulations that $\phi^*$ is equal to the angle of repose 
$\phi$.  Eq.~(\ref{stability1})
can be recast, using the invariants $P$ and $R$ into a stability condition of 
the form
\begin{equation}
Y\equiv \frac{R}{P\sin \phi} \le 1 \ ,
\label{stability2}
\end{equation}
everywhere in the sandpile. Specifically on the surface of the pile we expect 
marginal stability such that
the inequalities (\ref{stability1}) and (\ref{stability2}) are saturated.

Although Eqs.~(\ref{equil}) are under-determined, they do allow seeking scaling 
solutions in the form
\begin{eqnarray}
\sigma_{xx}&=& \frac{mgz}{\lambda^2} s_{xx}(S)\ , \nonumber\\
\sigma_{xz}&= &\sigma_{zx} = \frac{mgz}{\lambda^2} s_{xz}(S)\ , \nonumber\\
\sigma_{zz}&=& \frac{mgz}{\lambda^2} s_{zz}(S) \ ,
\label{scale}
\end{eqnarray}
where $S\equiv x\tan \phi/z$. Plugging this ansatz into Eqs.~(\ref{equil}) 
results in the scaled equations
\begin{eqnarray}
\tan\!\phi ~s'_{xx}+s_{xz} -S s'_{xz} &=& 0  \ , \nonumber\\
\tan\!\phi ~s'_{xz}+s_{zz} -S s'_{zz} &=& 1 .
\label{final}
\end{eqnarray}
where $s'\equiv ds/dS$.

To close these equations one needs a constitutive relation. Using 
Eq.~(\ref{defpsi}), dividing $\sigma_{xx}$ by
$\sigma_{xz}$ and rearranging we can derive the following relation
\begin{equation}
s_{xx}(S) =s_{zz} (S) -2 \cot \big(2\Psi(S)\big) s_{zx} (S)  \ .
\label{constit}
\end{equation}
This equation is of course useless as long as we do not know $\Psi(S)$. In 
general we are not
even guaranteed that $\Psi(S)$ exists as function of one variable.  In the 
general case we should
use instead of $\Psi(S)$ a function of two variables $\Psi(S,z)$. The dependence 
on one variable $S$ means that as the pile grows $S$ is changing and with it so 
does $\Psi(S)$. The strategy is therefore to determine the function $\Psi(S)$ 
from the numerics, and then to use Eq.~(\ref{constit}) as the constitutive 
relation that will
close the problem and will allow us to determine the stress field everywhere in 
the pile.

\subsection{Numerical Integration of the Scaling Solutions}

Eqs.~(\ref{final}) and (\ref{constit}) together with the boundary conditions 
$s_{xz}(S=1)=0$ and $s_{zz}(S=1)=0$ admit solutions
close to the free surface $S=1$ that are linear in $(1-S)$. These linear 
solutions as $S\to 1$ demand that $\cot(2\Psi(S=1))=\tan\phi$ or, equivalently 
\cite{95BCC,96WCCB,97WCC}, that
\begin{equation}
\Psi(S=1) = \frac{\pi -2\phi}{4} \ .
\label{psi1}
\end{equation}
There is less information about $\Psi(S=0)$ but simulations suggest (see below) 
that $\Psi(S=0)=0$. In other
words the principal axis ends up pointing in the direction of the gravity field 
as $S=0$ under the apex of the sandpile.

\begin{figure}
\includegraphics[width=0.45\textwidth]{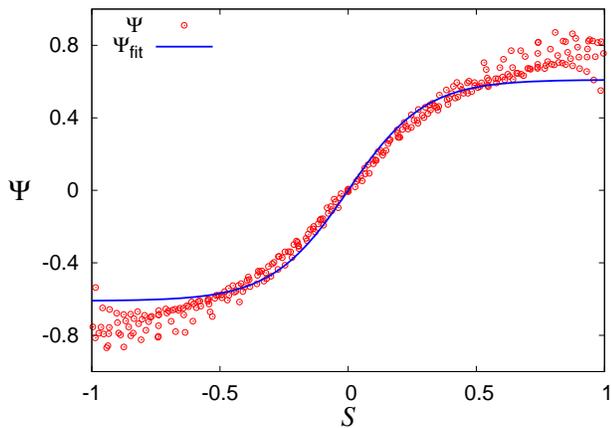}
\caption{The fitting function $\Psi_{\rm fit} (S)$ compared to the actual data 
for
$\mu=1$. }
\label{fit}
\end{figure}

As a first attempt in solving the problem we will use a fit to the numerical 
data that satisfies
the boundary condition (\ref{psi1}). We will see that this solution does not 
predict a dip in the pressure
at $S=0$. Examining our numerical estimate of $\Psi(S)$, see Fig.~\ref{fit},
%%%%%%%%%%%%%%%%%%%%%%%%%%%%%%%%%%%%%%%%%%%
%%%%%%%%%%%%%%%%%%%%%%%%%
we choose a global parameterized fit of the form
\begin{equation}
\label{psifit}
\Psi_{\rm fit} (S) = \frac{(\pi -2\phi)}{4} \frac{\tanh (\beta S)}{\tanh \beta} 
\ .
\end{equation}
We find that with $\beta= 3$ (cf. Fig.~\ref{fit}) we get a reasonable fit to the 
data that obeys
Eq.~(\ref{psi1}). A consequence of the agreement of our parameterized $\Psi_{\rm 
fit} (S)$  with the theoretical limit $\Psi(S=1)=(\pi-2\phi)/4$ as in 
Eq.~(\ref{psifit}) is that
the solution becomes marginally stable on the surface of the pile. This is 
physically pleasing since one expects
the outer surface to be just marginally stable and ready to avalanche with every 
addition of a new particle.
If we allowed the function $\Psi_{\rm fit} (S)$ to asymptote to the correct 
limit found in the simulations, which is higher than the expected theoretical 
value, it would lead to the loss of marginal stability at the surface. The 
solutions would become either stable or unstable at the surface
of the sandpile.

\begin{figure}
\includegraphics[width=0.40\textwidth]{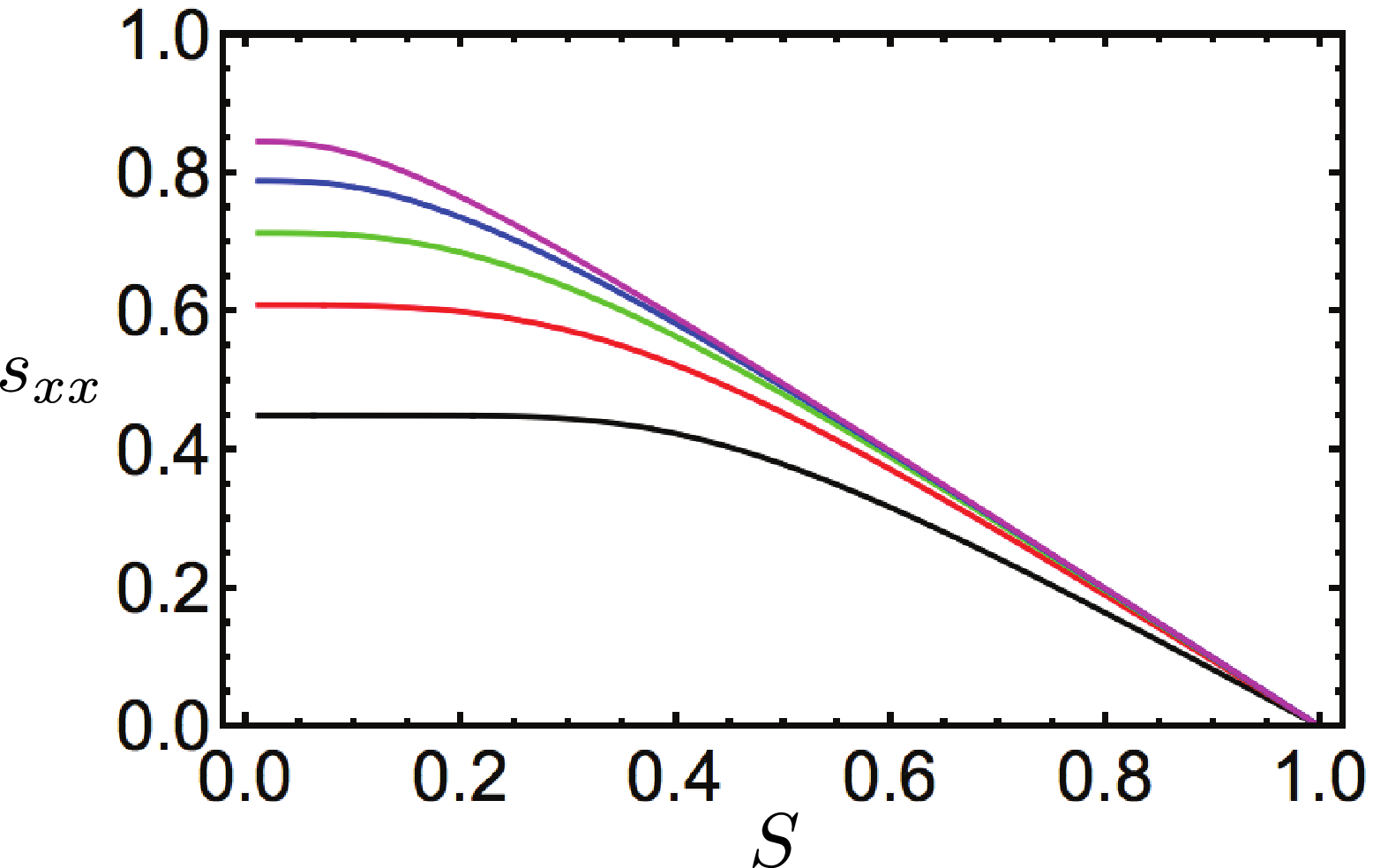}
\includegraphics[width=0.40\textwidth]{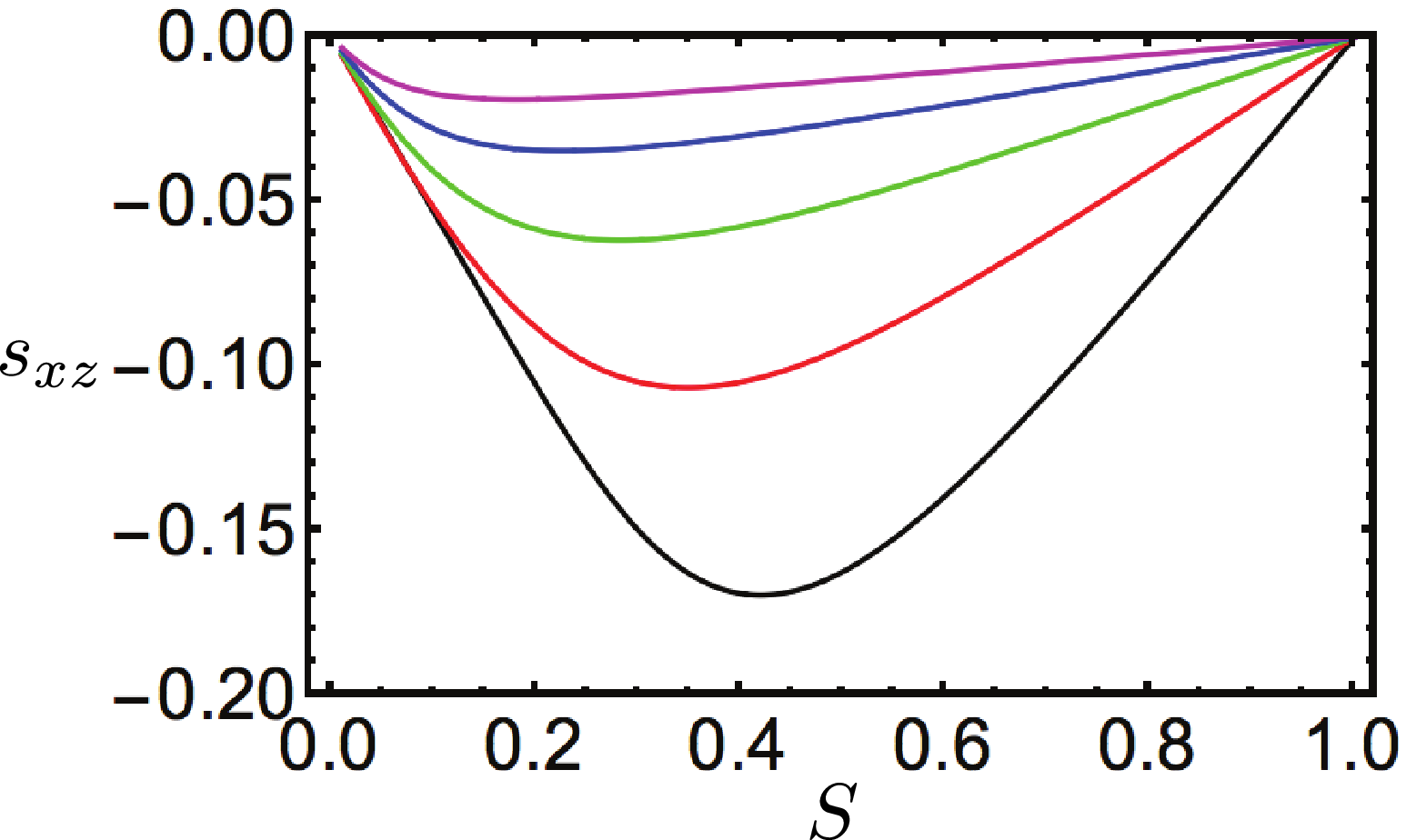}
\includegraphics[width=0.40\textwidth]{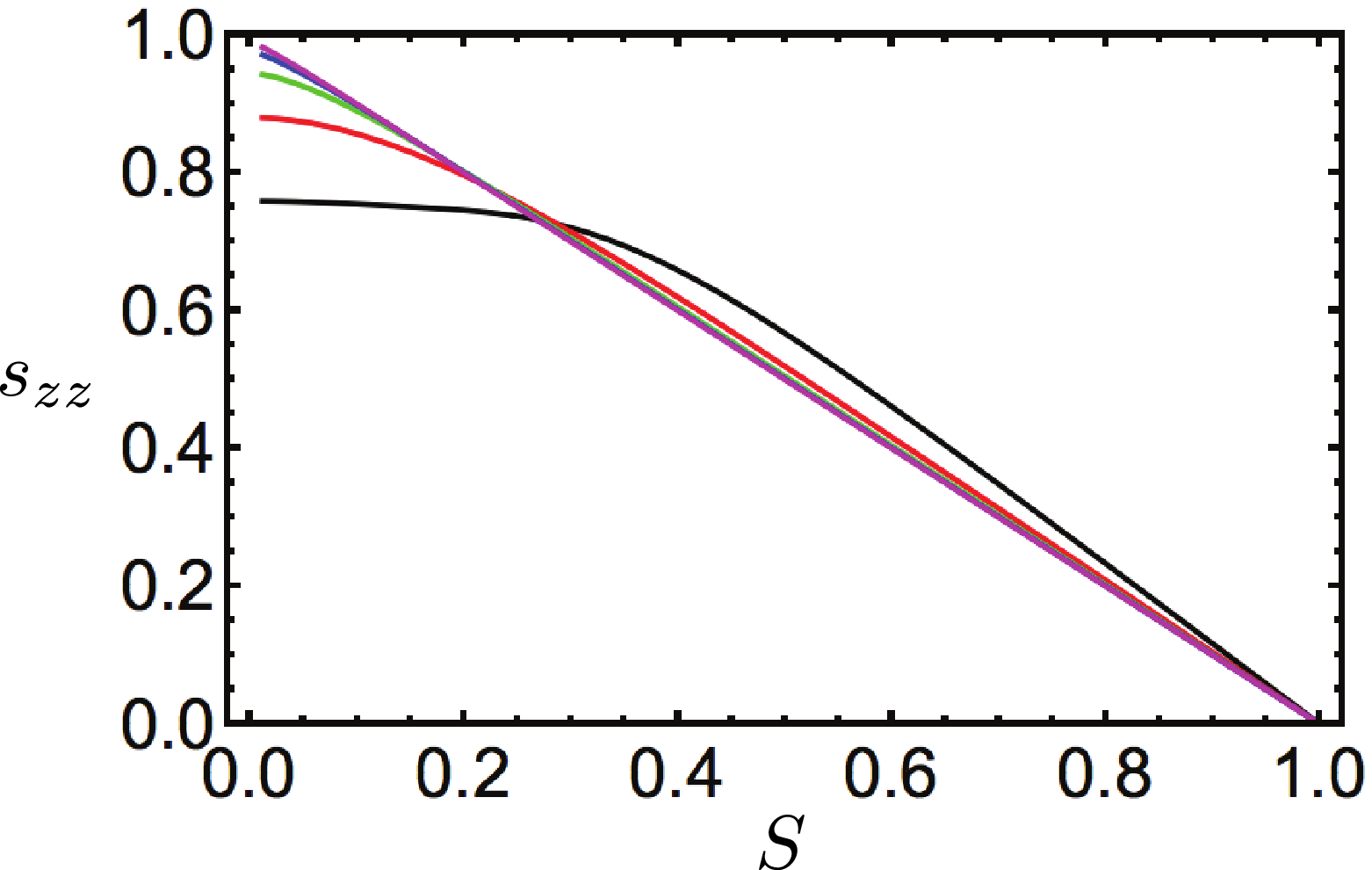}
\caption{The numerical results for the functions
$s_{xx}(S)$,  $s_{xz}(S)$ and  $s_{zz}(S)$ for sandpiles with different angle of 
repose found by solving Eqs.~\ref{final} inwards from $S=1$ where the boundary 
conditions are known.}
\label{numsol}
\end{figure}

The ordinary differential equations that should be solved are derived in the 
next subsection. Here we show that the strategy of integrating them from the 
outer boundary does not yield acceptable results. Starting from the given 
boundary condition we can solve numerically the linear coupled inhomogeneous 
ode's (\ref{ode1}) starting at the surface of the sand pile and integrating 
inward. In Fig.~\ref{numsol} we present solution for the three independent 
components of the stress tensor for different values of the angle of repose.
%%%%%%%%%%%%%%%%%%%%%%%%%%%%%%%%%%%%%%%%%%%
%%%%%%%%%%%%%%%%%%%%%%%%%%%%%%%%%%%%%%%%%%%%%%%%%%%%%%%%%%%%%%
The result is disappointing, there is no tendency to form a dip in any of the 
stress
components at the center of the pile. Should we believe these solutions?

\begin{figure}
\includegraphics[width=0.5\textwidth]{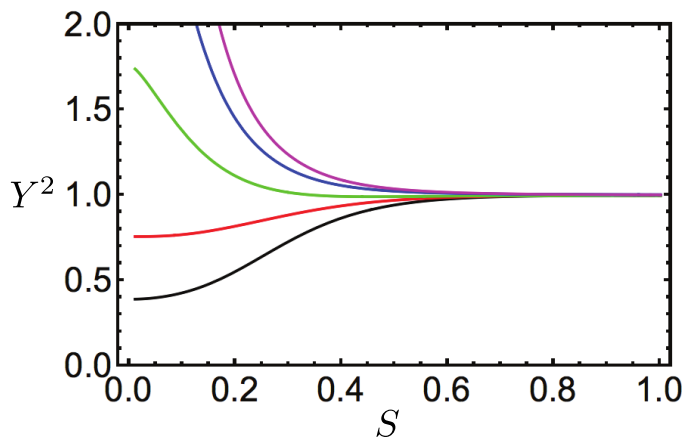}
\caption{The test of stability of the scaling solutions using $Y^2(S)$ for the 
same values of $\phi$
as in Fig.~\ref{numsol}.}
\label{stab}
\end{figure}

To see that these results are in fact untenable we should examine  the stability 
of these continuum solutions. To this aim we return to Eq.~(\ref{stability2}) 
and plot in Fig.~\ref{stab} the value of
$Y^2(S)$ for different values of the angle of repose.
%%%%%%%%%%%%%%%%%%%%%%%%%%%%%%%%%
%%%%%%%%%%%%%%%%%%%%%%%%%%%%%%%%%%%%%%%%%%%%%%%%%%%%%%%
We immediately note that deep in the sandpile, exactly where the peak to dip 
transitions can be expected to occur these scaling solutions become unstable, 
cf. Eq.~(\ref{stability2}).

 We conclude that the scaling solutions become unstable deep in the sandpile 
(see Fig.~\ref{stab}), especially near the center of the sandpile. It is just 
here, however, that the very interesting transitions in the pressure at the 
bottom of the sandpile from showing a peak to showing a dip appear. We
thus need to forsake the straightforward scaling solutions and examine the 
situation with greater scrutiny.

\section{Computer assisted theory}
\label{theory}

\subsection{Numerical evidence for the breaking of scaling}

\begin{figure}
\includegraphics[width=0.38\textwidth]{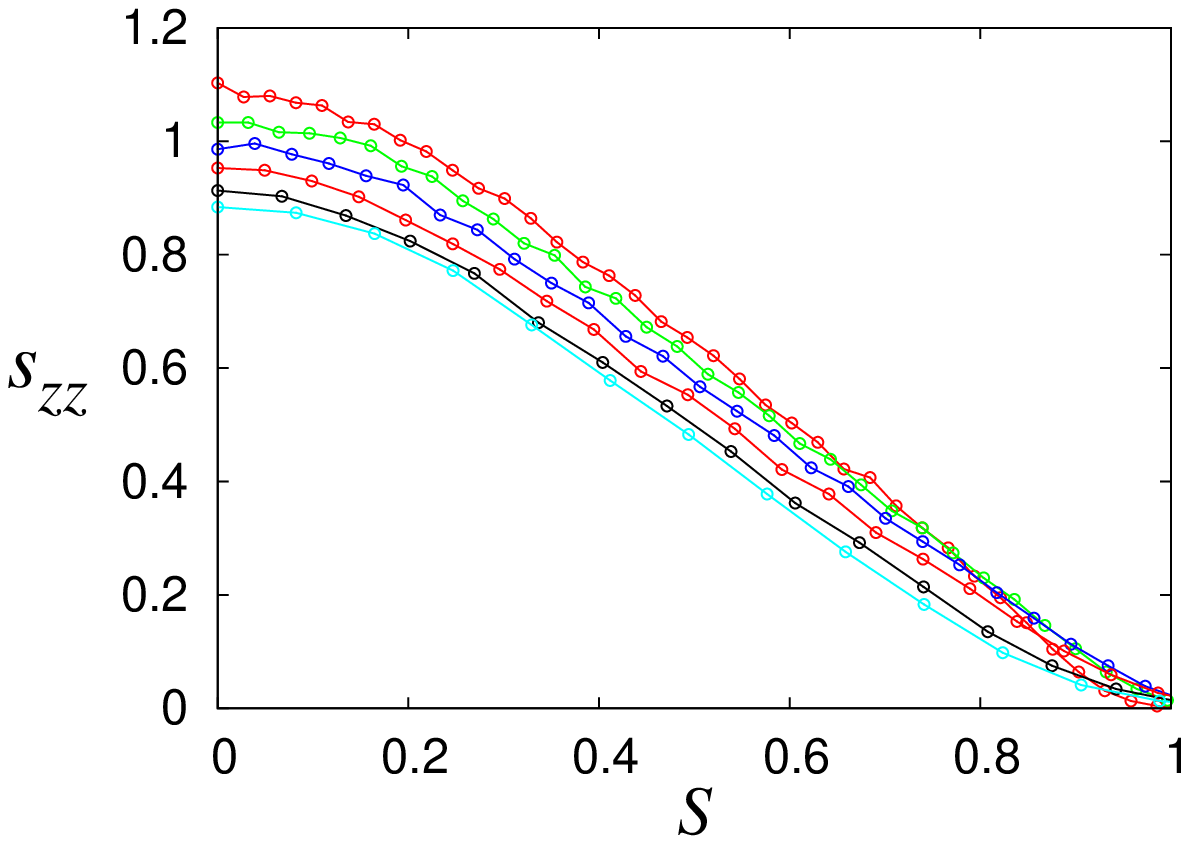}
\includegraphics[width=0.38\textwidth]{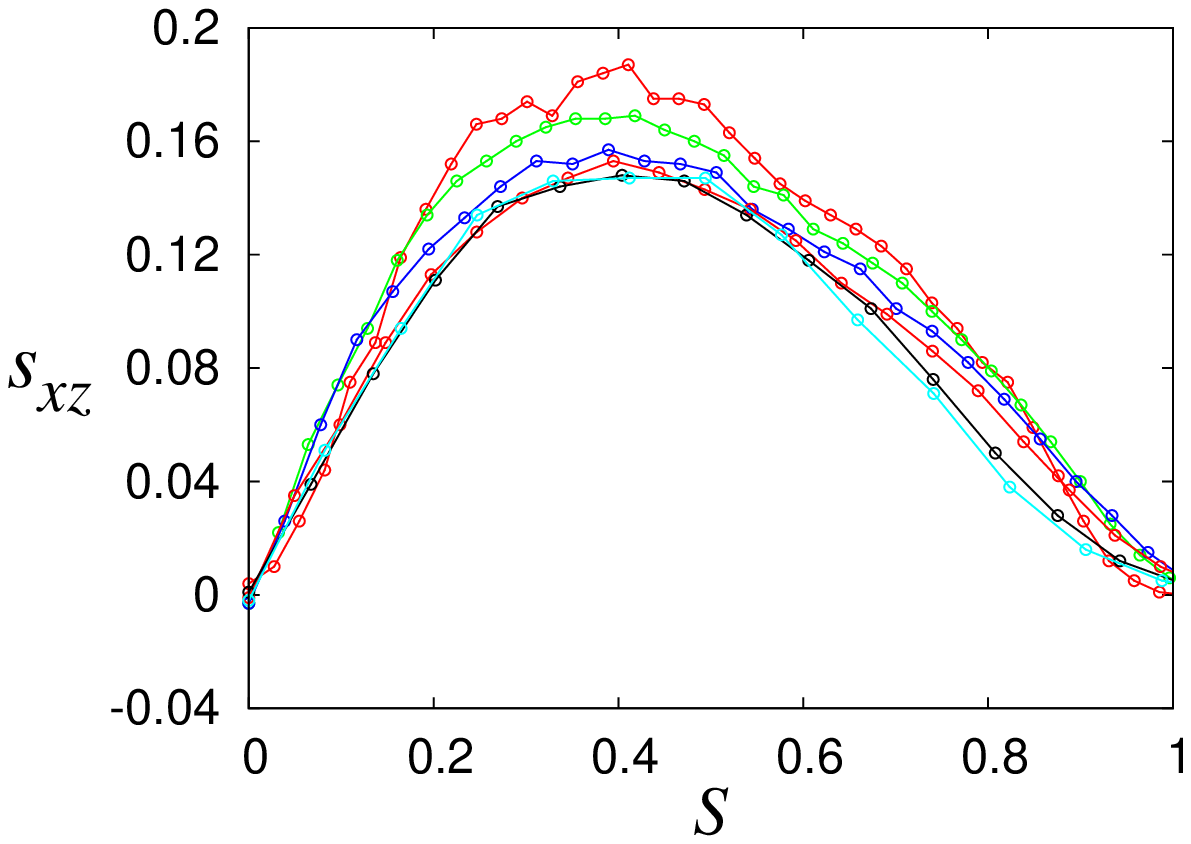}
\includegraphics[width=0.38\textwidth]{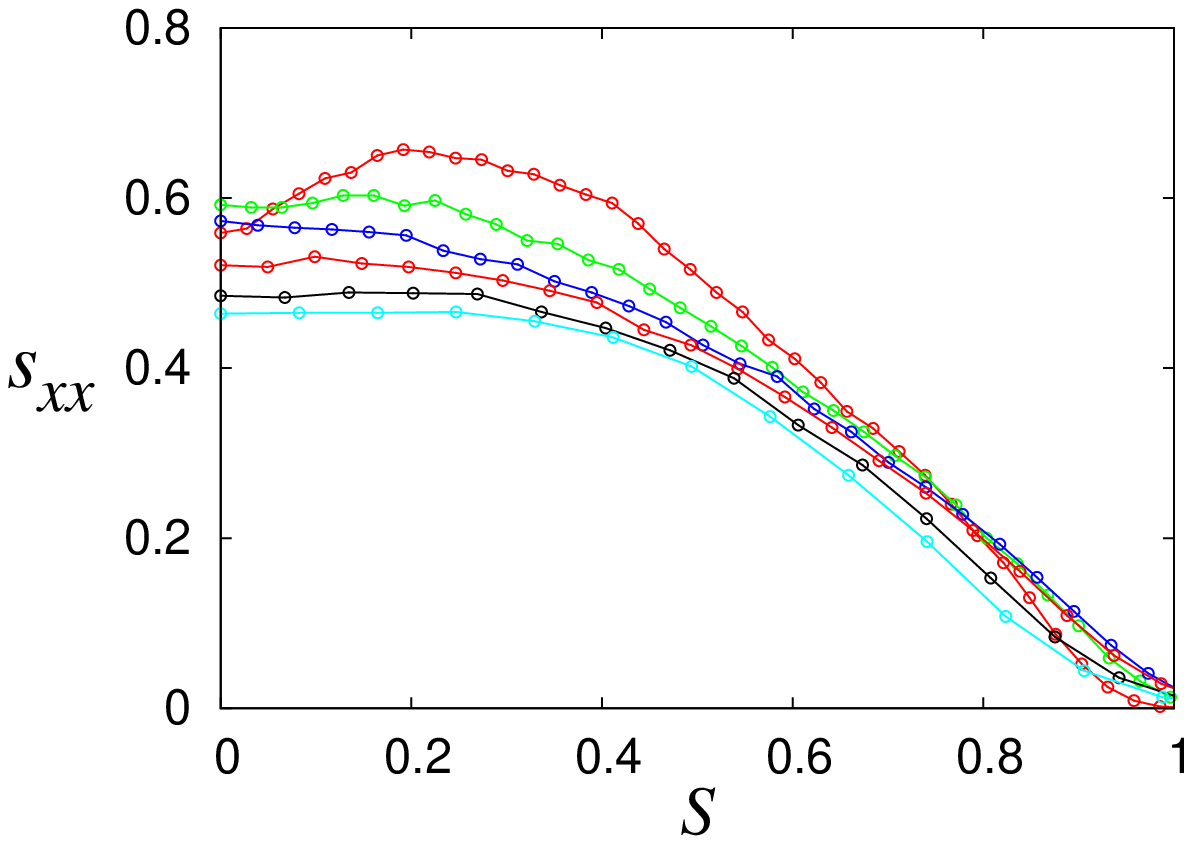}
\caption{The simulation results for the functions
$s_{zz}(S,z)$,  $s_{xz}(S,z)$ and  $s_{xx}(S,z)$ for one angle of repose, 
$\phi\approx \pi/8.73$ which pertains to $\mu=1$.
The different curves are now due to the fact that the scaling assumption is 
broken, and the functions are
shown at different heights $z$. The reader should not confuse these figures with 
Fig.~\ref{numsol} where the different color represent different angles of 
repose. The color code here refers to the following values of $z$ as
 measured from the apex: red : $z = 32.35$, green: $z = 27.55$, blue: $z = 
22.75$, magenta:  $z = 17.95$, black: $z = 13.15$, cyan: $z = 8.35$.}
\label{simsol}
 \end{figure}

Having extracted the
stress tensor in the pile as a function of $(x,z)$ we can compute the 
theoretically relevant tensors
$s_{xx}(S,z)$, $s_{xz}(S,z)$ and $s_{zz}(S,z)$. These are plotted for $\mu=1$ 
and $\phi\approx0.36= \pi/8.73$ for different values of $z$, in the three panels 
of Fig.~\ref{simsol}. In comparing the plots in Fig.~\ref{numsol} and 
Fig.~\ref{simsol} one should observe that the model
simulations are not precisely scaling and depend explicitly on $z$. This should
not be confused with the theoretical curves in Fig.~\ref{numsol} that pertain to 
different angles of repose but that show perfect scaling.
%%%%%%%%%%%%%%%%%%%%%%%%%
%%%%%%%%%%%%%%%%%%%%%%%%%%%%%%%%%%%%%%%%%%

Clearly, scaling appears to be broken and the "scaling" solutions $s_{xx}(S)$,  
$s_{xz}(S)$ and  $s_{zz}(S)$  are actually functions of both $(S,z)$. The same 
problem exists with the function $\Psi(S)$ as discussed in the
next subsection.

\subsection{The Direction Of the Principal Stress Axis}
Solving Eqs.~(\ref{final}) is impossible without adding an explicit form for the 
principal stress axis $\Psi$ into the constitutive relation Eq.~(\ref{constit}) 
to close the mathematical problem.
A number of constitutive relations have been proposed; we find the so-called 
``Fixed Principal Axis" quite plausible,
since it is based on the physical assumption that once the principal axis of the 
stress was determined near
the surface, it gets buried unchanged with the growth of the sandpile. Note that 
this makes no assumption
about the {\em magnitude} of the principal stresses, only about their 
directions. Nevertheless our
simulations are not in full agreement with this assumption, neither near the 
surface nor deep in the pile.
We therefore turn now to the analysis of the principal axis of the stress field 
as seen in simulations.

%%%%%%%%%%%%%%%%%%%%%%%%%%%%%%%%%%%%%%%%%%
\begin{figure}
\vskip -0.5 cm
\includegraphics[width=0.38\textwidth]{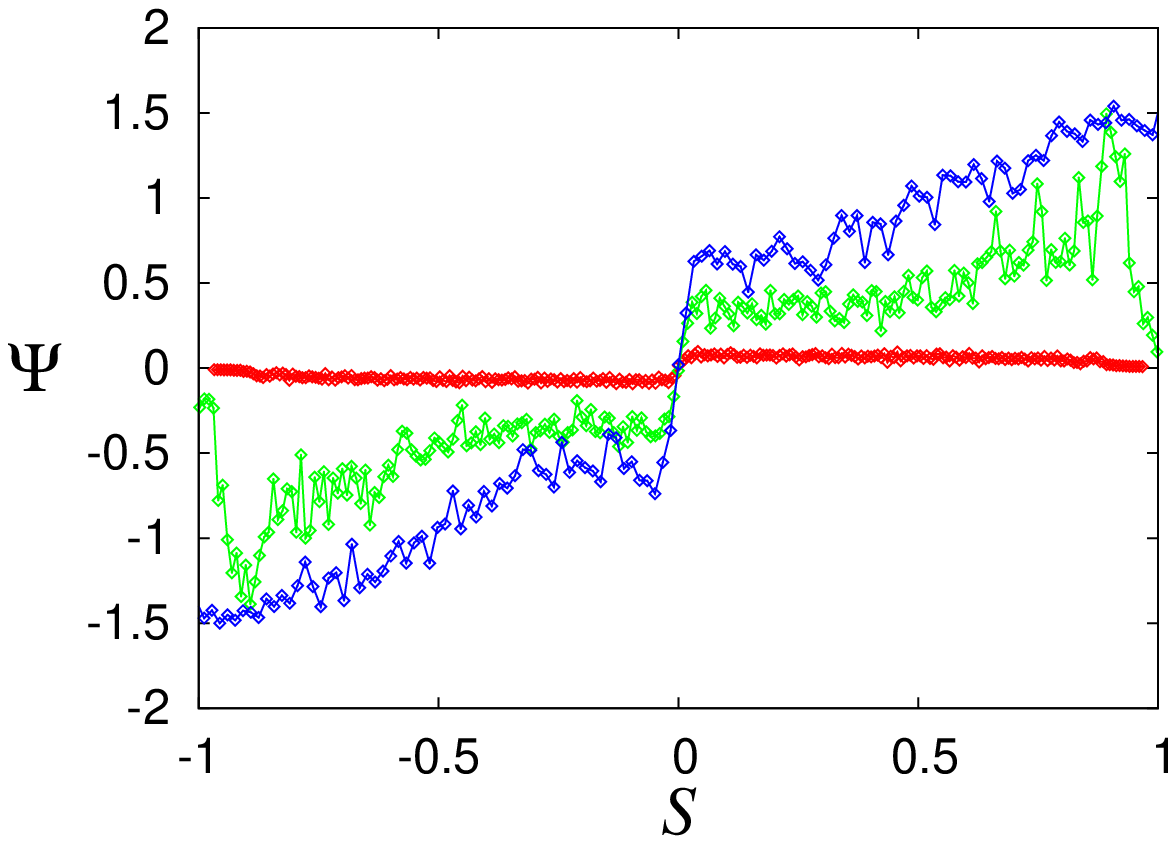}
\includegraphics[width=0.38\textwidth]{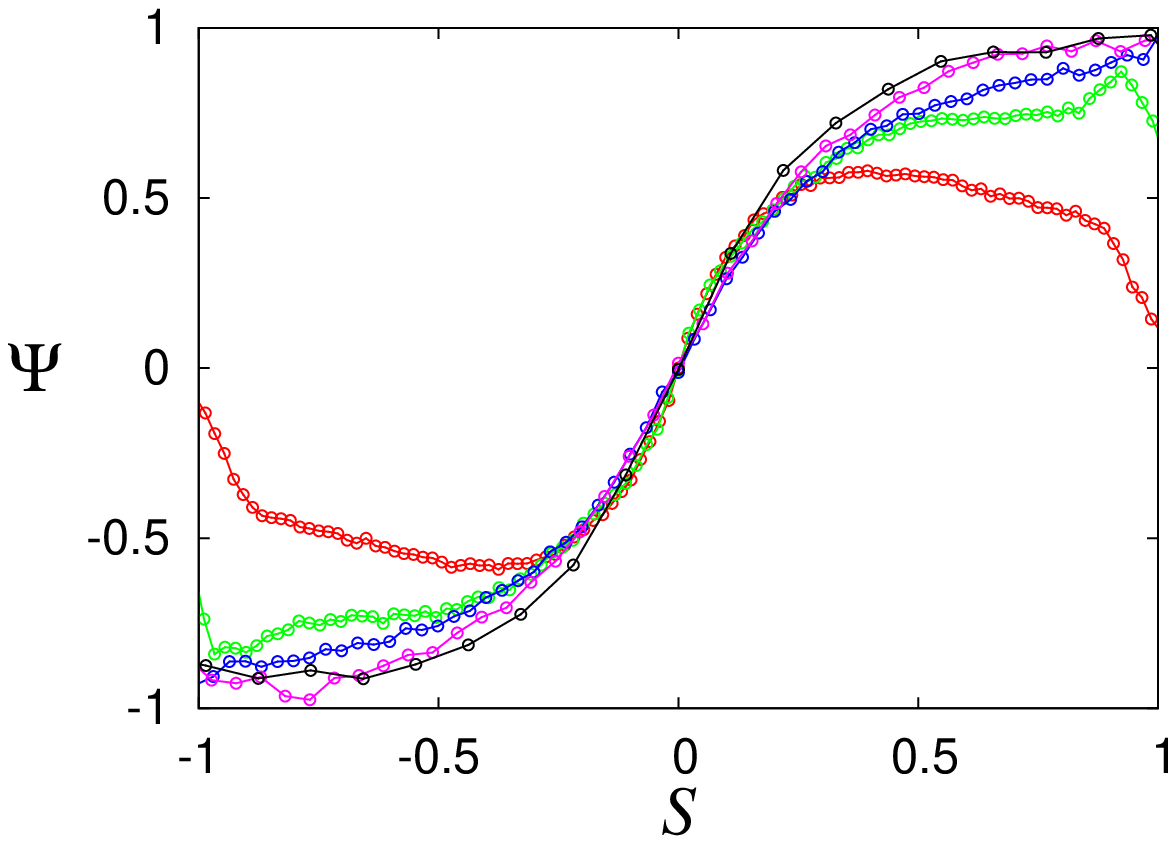}
\includegraphics[width=0.38\textwidth]{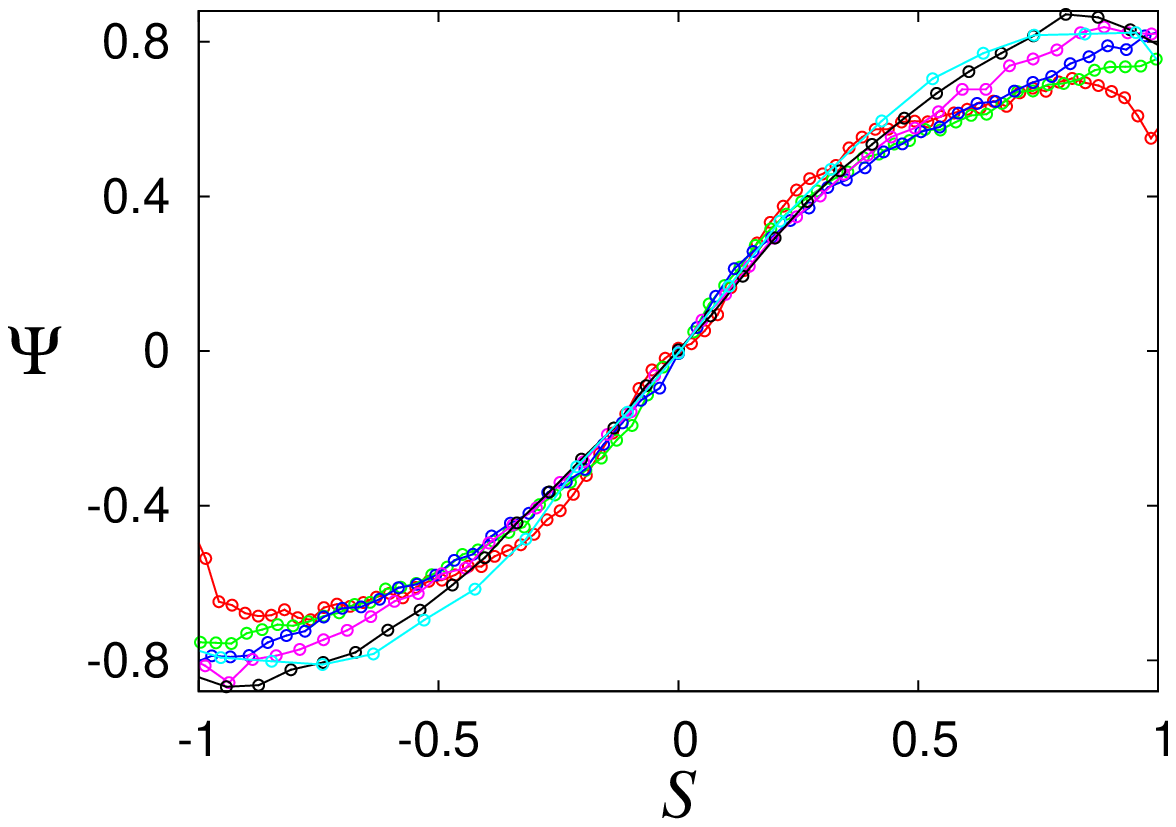}
\caption{The angle $\Psi(S,z)$ as a function of $S$ computed for sandpiles 
created with $\mu=0.01$ (upper panel), $\mu=0.1$ (middle panel) and $\mu=1$ 
(lower panel). The different colors correspond to different heights $z$.
 If the scaling assumption prevailed all these plots should collapse. Note that 
for $\mu=1$ the scaling assumption is
 close to being obeyed but it breaks totally for $\mu=0.01$. The color code 
refers to different values of $z$ as
 measured from the apex which are: upper panel; red : $z = 8.28$, green: $z = 
5.88$, blue: $z = 3.48$. Middle panel;
 red : $z = 23.42$, green: $z = 18.62$, blue: $z = 13.82$, magenta:  $z = 9.02$, 
black: $z = 4.22$.
 Lower panel; red : $z = 32.35$, green: $z = 27.55$, blue: $z = 22.75$, magenta: 
 $z = 17.95$, black: $z = 13.15$, cyan: $z = 8.35$.}
\label{psilines}
\end{figure}
%%%%%%%%%%%%%%%%%%%%%%%%%%%%%%%%%%%%%%%%%%
 When the scaling assumptions discussed above are valid, $\Psi$
should be a function of $S$ only, $\Psi(S)$. If scaling breaks down this 
function depends on both $S$ and $z$. The numerical
 evidence is shown in Fig.~\ref{psilines}. The three panels present $\Psi(S,z)$ 
at different heights $z$ in the sandpile. If the scaling assumption were 
corroborated by the simulation we would expect that all
these curves would collapse on a single function $\Psi(S)$. We learn that for 
large values of the friction coefficient
$\mu$, which correspond to large angle of repose, the scaling assumption is 
reasonably obeyed. The scaling assumption
deteriorates when $\mu$ and the angle of repose decrease, until eventually it 
breaks down entirely. Notice however that the actual function $\Psi(S)$, {\em 
even when the scaling assumption is relatively acceptable} does not agree with 
many
constitutive relations that were assumed in the literature, which are often 
piece-wise linear, cf. Refs.~\cite{95BCC,96WCCB,97WCC}
and reference therein.

For large of values of $\mu$, $\Psi(S,z)$ does indeed approach a scaling 
function $\Psi(S)$ except for coordinates at the
bottom of the pile, cf. the curve shown in red in lower panel of 
Fig.~\ref{psilines}. Note that even in those cases $\lim_{S\to 1} \Psi (S)$ is 
not very close the theoretical expectation Eq.~(\ref{psi1}), which for $\mu=1$ 
and $\phi=0.36$ predicts $\Psi(S=1)\approx 0.61$. We propose that this 
discrepancy
stems from the large fluctuations in the surface geometry which manifests itself 
in the large fluctuations in $\Psi$ seen in
Fig.~\ref{psidir} at the surface, and see below for more details.
%%%%%%%%%%%%%%%%%%%%%%%%%%%%%%%%%%%%%%%%%%%%%%%%%%%%%%%%%%%%%%%%%%
\subsection{Analytic solutions Valid Near The Center Of The Sandpile}

When we integrate  the scaled equations (\ref{final})
from the surface of the sandpile using the boundary conditions $s_{xz}(S=1)=0$ 
and $s_{zz}(S=1)=0$
the solutions become unstable deep in the sandpile. We therefore change our 
approach here and attempt to  supplement our numerical integration of 
Eqs.~(\ref{final}) and Eq.~(\ref{constit}) with analytical approaches valid deep 
in the sandpile. Thus
substituting Eq.~(\ref{constit}) in Eqs.~(\ref{final}) and rearranging
we find  for
\begin{equation}
\label{s}
 {\B s}(S) = \left(
\begin{array}{c}
s_{xz}(S)\\
s_{zz}(S) \\
\end{array}
\right).
\end{equation}
the set of coupled linear inhomogeneous ordinary differential equations
\begin{equation}
\label{ode1}
\B A(S) \B s'(S) +\B D(S) \B s(S) = \left(
\begin{array}{c}
0\\
1 \\
\end{array}
\right).
\end{equation}
Here
\begin{equation}
\B A(S) \equiv \left(
\begin{array}{cc}
-\Big(2\tan\phi \cot \big(2 \Psi(S)\big) +S\Big) & \tan\phi \\
\tan \phi & -S \\
\end{array}
\right).
\end{equation}
and
\begin{equation}
\B D(S) \equiv \left(
\begin{array}{cc}
1+\frac{4\tan\phi}{\sin^2 \big(2\Psi(S)\big)}\Psi'(S) & 0 \\
0 & 1 \\
\end{array}
\right).
\label{odemat}
\end{equation}

We cannot simply integrate from the center of the sandpile ($S=0$) towards the 
surface, however, as we do not possess the required boundary value for 
$s_{zz}(0)$. We will therefore take another approach.

Eq.~(\ref{ode1}) is in principal exactly soluble as it is a linear inhomogenous 
equation, and we also have a good knowledge of the symmetry properties of all 
the relevant variables near the center of the sandpile.  We can therefore find a 
solution that is valid order by order in $S$ for small $S$ near the center of 
the sandpile.

First we will write the general solution of  Eqs~(\ref{ode1}) as the sum of a 
particular solution and a complementary solution of the associated homogeneous 
equation as $\B s(S) = \B s_p(S) + \B s_{hom}(S)$ where the particular solution 
is
\begin{equation}
\label{sp}
 \B s_p(S) = \left(
\begin{array}{c}
0\\
1\\
\end{array}
\right).
\end{equation}
and the complementary solution obeys (using from now on the notation $t\equiv 
\tan \phi$)
\begin{equation}
d\B s_{hom}(S)/dS +\B B(t, S) \B s_{hom}(S) = \left(
\begin{array}{c}
0\\
0 \\
\end{array}
\right).
\label{hom}
\end{equation}
Here $\B B = \B A^{-1} \B D$ or
\begin{eqnarray}
&& \B B(t, S) \equiv -{\det \B A}^{-1}  \\
&& \left(
\begin{array}{cc}
S \big(1+\frac{4 t}{\sin^2 \big(2\Psi(S)\big)}  \Psi'(S)\big ) & t\\
t \big(1+\frac{4 t}{\sin^2 \big(2\Psi(S)\big)}  \Psi'(S)\big ) &\Big(2 t \cot 
\big(2 \Psi(S)\big) +S\Big)
\end{array}
\right)\nonumber \ .
\label{ode2}
\end{eqnarray}
where $\det \B A(t,S) = S (2 t \cot (2 \Psi(S)\big) +S) - t^2$.

Eq.~(\ref{hom}) can be integrated exactly from $S=0$ to $S=1$ and using the 
boundary outer boundary conditions
\begin{equation}
\label{uout}
 \B s_{hom}(1) = \left(
\begin{array}{c}
0\\
-1\\
\end{array}
\right).
 \end{equation}
we can find an explicit equation for $\B s(S)$ in the form
\begin{equation}
\label{ss}
 \B s(t,S) = (I - \B M^{-1}(t,S) \B M(t,1) )\left(
\begin{array}{c}
0\\
1\\
\end{array}
\right).
\end{equation}
where
\begin{equation}
\label{M}
\B M(t,S) = \exp{\int_0^S \B B(t,S') dS'}.
\end{equation}
Eq.~\ref{M} is a $2\times2$ matrix that depends explicitly on the scaling 
variable $S$ and the angle of repose $\phi$ through the variable $t\equiv 
\tan{\phi}$. It is a global function of $S$. For example,
we can find an explicit expression for $s_{zz}(t,0)=1- M(t)$ at the center of 
the sand pile and consequently an expression for the pressure at the center of 
the sandpile in terms of a global integral over the whole sandpile
\begin{equation}
\label{M22}
M(t) = \big(\exp{\int_0^1 \B B(t,S') dS'}\big)_{22}.
\end{equation}

We are especially interested in the behavior of the scaling variables near the 
center of the sandpile (small $S$) as this will allow us to see any peak to dip 
transitions transitions in the stress. We shall now expand  the angle of the 
principal axis near the center of the sandpile as
\begin{equation}
\label{psiex}
\Psi(S) = \alpha S -\beta S^3 + \gamma S^5 +\cdots \ ,
\end{equation}
using the fact that this angle is an odd function of $S$ (see 
Fig.~\ref{psilines}). We will assume that $\alpha$ and $\beta$ are dimensionless 
parameters dependent on the material properties of the sandpile.

We are now in a position to calculate the series expansion of the complete 
stress tensor as a series expansion in $S$. Using the symmetry properties of the 
scaled stress tensor components we write
\begin{widetext}
\begin{eqnarray}
s_{xz}(t,\alpha,S) & = & s_{xz1}(t,\alpha)S + s_{xz3}(t,\alpha)S^3 + 
s_{xz5}(t,\alpha)S^5  + \cdots \nonumber \\
s_{zz}(t,\alpha, S) & = & s_{zz0}(t,\alpha) + s_{zz2}(t,\alpha)S^2 + 
s_{zz4}(t,\alpha)S^4 + s_{zz6}(t,\alpha)S^6 + \cdots \nonumber \\
s_{xx}(t,\alpha,S) & = & s_{xx0}(t,\alpha) + s_{xx2}(t,\alpha)S^2  + 
s_{xx4}(t,\alpha)S^4 + s_{xx6}(t,\alpha)S^6 + \cdots \nonumber \\
\label{st}
\end{eqnarray}
\end{widetext}
Substituting  expressions ~(\ref{st}) into the scaled continuum 
Eq.~(\ref{final}b), we can expand order by order in $S$ and in this manner get 
explicit expressions for the following coefficients of the stress tensor:
\begin{eqnarray}
\label{st1}
s_{xz1}(t) & = &  \frac{1-s_{zz0}}{t}\nonumber\\
s_{xz3}(t,\alpha) & = &  \frac{s_{zz2}}{3t} \nonumber \\
s_{xz5}(t,\alpha) & = &  \frac{3s_{zz4}}{5t}\nonumber \\
s_{xz7}(t,\alpha) & = &  \frac{5s_{zz6}}{7t} .
\end{eqnarray}
Now, substituting expressions ~(\ref{st}) into Eq.~(\ref{final}a), we can again 
expand order by order in $S$ and get the following relations:
\begin{eqnarray}
\label{st2}
s_{xx2}(t) & = &  0\nonumber\\
s_{xx4}(t,\alpha) & = &  \frac{s_{xz3}}{2t} = \frac{s_{zz2}}{6 t^2} \nonumber \\
s_{xx6}(t,\alpha) & = &  \frac{2s_{xz5}}{3t} = \frac{2 s_{zz4}}{5 t^2}\nonumber 
\\
s_{xz7}(t,\alpha) & = &  0 ,
\end{eqnarray}
Finally, we substitute expressions ~(\ref{st}) into the constitutive equation 
Eq.~(\ref{constit}) and obtain the following expressions:
\begin{eqnarray}
\label{st3}
s_{xx0}(t,\alpha) & = &  s_{zz0}-\frac{s_{xz1}}{\alpha}\nonumber\\
s_{xx2}(t,\alpha) & = &  s_{zz2}-\frac{s_{xz3}}{\alpha}+\frac{4\alpha 
s_{xz1}}{3}-\frac{\beta s_{xz1}}{\alpha^2}\nonumber \\
s_{xx4}(t,\alpha) & = &   s_{zz4}-\frac{s_{xz5}}{\alpha}+\frac{4\alpha 
s_{xz3}}{3}-\frac{\beta s_{xz3}}{\alpha^2}
+\frac{16\alpha^3 s_{xz1}}{45}\nonumber \\
&-&\frac{4\beta s_{xz1}}{3}-\frac{\beta^2 s_{xz1}}{\alpha^3}+\frac{\gamma 
s_{xz1}}{\alpha^2}\nonumber \\
s_{xx6}(t,\alpha) & = &  s_{zz6}+\frac{128 \alpha ^5 s_{xz1}}{945}-\frac{\beta 
^3 s_{xz1}}{\alpha ^4}-
\frac{16 \alpha ^2 \beta  s_{xz1}}{15}\nonumber \\
&+& \frac{4\gamma s_{xz1}}{3} + \frac{2\beta\gamma s_{xz1}}{\alpha^3} - 
\frac{\beta ^2 s_{xz3}}{\alpha ^3}+\frac{16 \alpha ^3 s_{xz3}}{45}\nonumber \\
&-&\frac{4 \beta  s_{xz3}}{3}+\frac{\gamma s_{xz3}}{\alpha ^2}-\frac{\beta  
s_{xz5}}{\alpha ^2} + \frac{4 \alpha  s_{xz5}}{3} .
\end{eqnarray}
Using $s_{zz0}(t) = 1-M(t)$, we can further simplify the above expressions to 
obtain
\begin{eqnarray}
\label{st0}
s_{zz0}(t) & = &  1-M(t)\nonumber\\
s_{xz1}(t) & = & M(t)/t \nonumber \\
s_{xx0}(t,\alpha) & = & 1-M(t)(1+\frac{1}{\alpha t}) \nonumber \\
s_{zz2}(t,\alpha) & = &  \frac{M(t)(4 \alpha^3 - 3 \beta)}{\alpha (1 - 3 t 
\alpha)}) \nonumber\\
s_{xz3}(t,\alpha) & = &\frac{M(t)(4 \alpha^3-3\beta)}{3 t \alpha (1-3 t 
\alpha)}\nonumber \\
s_{xx2}(t,\alpha) & = & 0 \ ,
\end{eqnarray}
\begin{widetext}
\begin{eqnarray}
s_{zz4}(t,\alpha) & = & \frac{M(t)(-4 \alpha^4(5+8 t \alpha(-2+t \alpha))+ 15 
\alpha(1+8 t \alpha(-1+t \alpha))\beta+90 t^2 \beta^2+30t\gamma (1-3t\alpha))}{ 
6 t \alpha (-1+3 t \alpha)(-3+5 t \alpha)} \nonumber \\
s_{xz5}(t,\alpha) & = &\frac{M(t)(-4 \alpha^4(5+8 t \alpha(-2+t \alpha))+15 
\alpha(1+8 t \alpha (-1 +t \alpha))\beta + 90 t^2 \beta^2+30t\gamma 
(1-3t\alpha))}{10 t^2 \alpha(-1+3 t \alpha )(-3+5 t \alpha ) } \nonumber \\
s_{xx4}(t,\alpha) & = & \frac{M(t)(4 \alpha^3- 3 \beta)}{6 t^2 \alpha( 1 - 3 t 
\alpha)} \nonumber \\
s_{zz6}(t,\alpha) & = & 0 \nonumber \\
s_{xx6}(t,\alpha) & = & \frac{M(t)(-4 \alpha^4(5+8 t \alpha(-2+t \alpha))+ 15 
\alpha(1+8 t \alpha(-1+t \alpha))\beta+90 t^2 \beta^2+30t\gamma (1-3t\alpha))}{ 
15 t^3 \alpha (-1+3 t \alpha)(-3+5 t \alpha)}
\end{eqnarray}
\end{widetext}
Let us analyze $s_{zz2}(t,\alpha)  =   [M(t)/\alpha)(4 \alpha^3 - 3 \beta)/(1 - 
3 t \alpha)]$ more closely.  We first note that if  $(4 \alpha^3- 3 \beta  )=0$ 
the coefficient goes through zero. Thus a critical value of the principal axis 
gradient exists, given by
\begin{equation}
\label{crit}
\alpha_c = (3\beta/4)^{1/3} .
\end{equation}
This will create  the well known peak to dip transition in sand pile pressure at 
the center of the sandpile.
There is also a second transition apparent in the coefficients.
We can also see that at lower values of $\alpha$ a singularity develops in the 
coefficients when
$(1-3 t \alpha )=0$. Or
\begin{equation}
\label{sing}
\alpha_{sing} = \frac{1}{3t}.
\end{equation}

\begin{figure}
\includegraphics[width=0.32\textwidth]{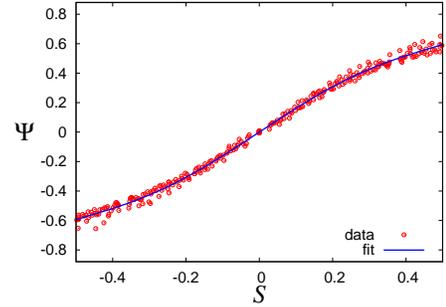}
\caption{A plot of principal stress direction $\Psi(S)$ near the center of the 
pile. Red circles are the raw data obtained from the simulation and the blue 
curve is a fit (Eq.~(\ref{psiex})) to the data. The fitting parameters are 
$\alpha=1.64052, \beta=2.77268, \gamma=3.872$.}
\label{psifit2}
\end{figure}

\begin{figure}
\includegraphics[width=0.32\textwidth]{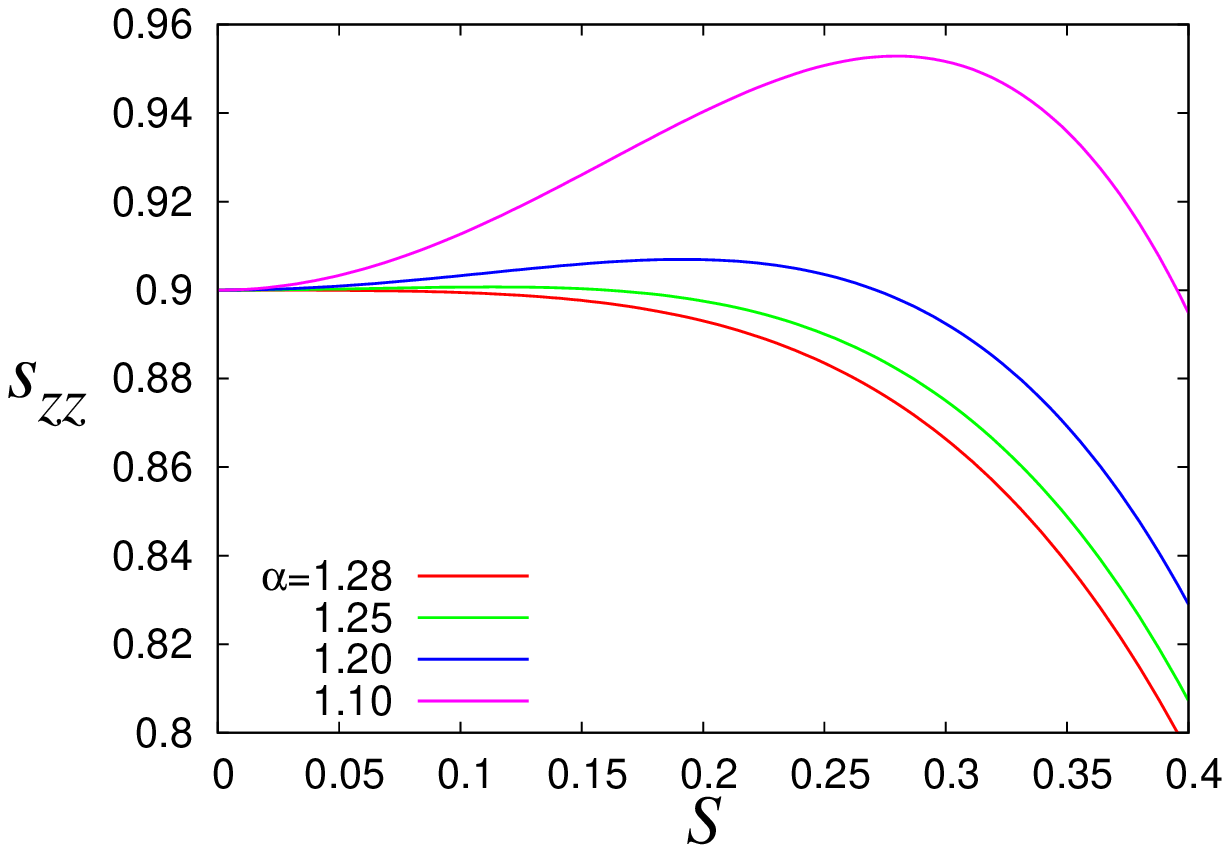}
\includegraphics[width=0.32\textwidth]{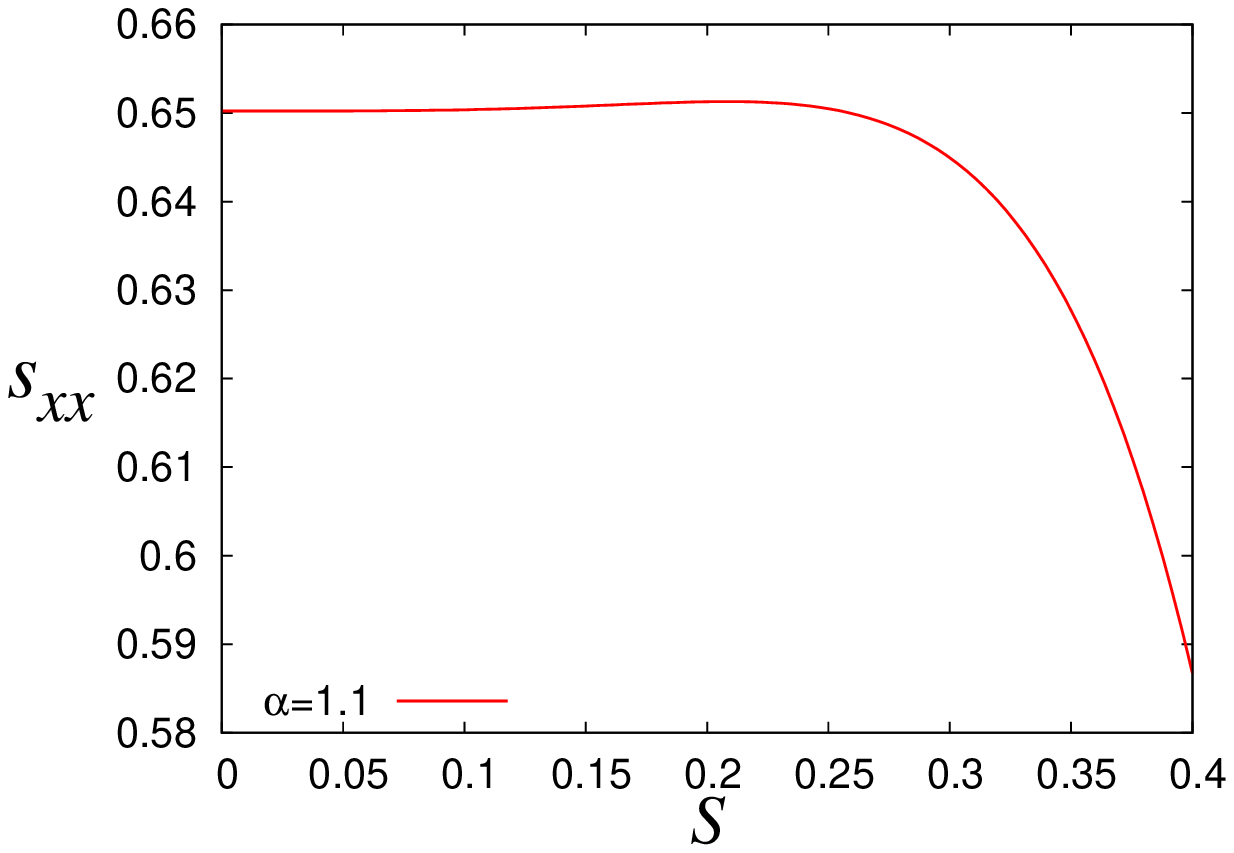}
\caption{Plots of $s_{zz}$ (upper panel) and $s_{xx}$ (lower panel) as a 
function of $S$ and $\alpha$ near the center of the sandpile. Note the 
transition from a peak to a dip as the principal axis gradient $\alpha$ 
decreases. We have chosen $\phi =\pi/9.0,\beta=2.77268$ and $\gamma=3.872$ for 
these plots.}
\label{szzxxfig}
\end{figure}

A fitting of principal stress direction $\Psi(S)$ near the center of the pile 
using Eq.~(\ref{psiex}) is shown in Fig.~\ref{psifit2}. Now, using these fitted 
values of $\beta$ and $\gamma$, we compute $s_{zz}(\alpha,S)$ and 
$s_{xx}(\alpha,S)$ for $t=\pi/9.0$, which corresponds to the angle of repose for 
$\mu=1$. These functions are shown in Fig.~\ref{szzxxfig}. We see that there is 
a transition in $s_{zz}(\alpha,S)$ from a peak ($s_{zz2} < 0$) to a dip 
($s_{zz2} > 0$) as $\alpha$ decreases.
This is in accordance with our simulation results as smaller values of $\alpha$ 
will correspond to a higher $\mu$. A similar but less prominent dip is also 
observed in $s_{xx}(\alpha,S)$ in the same range of $\alpha$.
Furthermore, we also find a singularity in the second and higher order 
coefficients at $\alpha=1/(3t)\approx 0.955$, but it is not clear at present 
whether this is an observable singularity. The reason being that 
Eq.~(\ref{crit}) depends on the parameter $\beta$ and for real materials this 
may lead to $\alpha_{crit}\gg \alpha_{sing}$ for sandpiles.
%%%%%%%%%%%%%%%%%%%%%%%%%%
%%%%%%%%%%%%%%%%%%%%%%%%%%
%%%%%%%%%%%%%%%%%%%%%%%%%%
%%%%%%%%%%%%%%%%%%%%%%%%%%

We can conclude that the theory proposed here is in agreement with the observed 
phenomenology
concerning the creation of a dip in the base pressure. What remains is
to consider some of the physical reasons why a breakdown in scaling may exist in 
sandpiles, and how this phenomenon is related to the value of $\mu$, or 
equivalently to the value of the angle of repose.

%%%%%%%%%%%%%%%%%%%%%%%%%%%%%%%%%%%%%%%%%%%%%%%%%%%%%%%%%%%%%%%%%%%%%%%%%%%%%%%%
%%
\section{New Length Scales}
\label{scaling}
It appears that there exists a clear transition at some critical angle of 
repose, say
$\phi_{\rm crit}$, below which {\em scaling solutions} stop being stable inside 
the sandpile. This does not mean that there cannot exist stable solutions to the 
mechanical equilibrium problem, but rather, as we see in the simulations, 
scaling
is broken and the solutions become heterogeneous, dependent on both $x$ and $z$. 
Thus in some sense the theory
predicts its own demise as is corroborated by the simulations. This is not a 
trivial remark. We have taken a ``reasonable"
constitutive relation fitted to the case $\mu=1$ where the scaling assumption 
seems to be approximately obeyed. We find that
this constitutive relation predicts its own irrelevance for lower values of 
$\mu$ as the predicted scaling solutions become
mechanically unstable. Here we explore several reasons why  broken scaling may 
be expected.

We need to stress at this point  that the whole formalism we have developed thus 
far can only be accepted when the gravity $g$ acting on cohesion-less particles 
provides the only length scale in the problem. There are, however,  at least 
three potential length scales that may interfere with these
assumptions: (i) there are elastic length scales which will involve the bulk and 
shear moduli, (ii) there
exist a bifurcation length scales that determines the distance between 
bifurcations in the force chains as
seen in the simulations, and (iii) a length scale associated with surface 
roughening or height fluctuations
on the surface due to ever-existing avalanches. Here we examine how this 
modifies
the  theory presented above.

\subsection{The effect of surface fluctuations}
\begin{figure}
\includegraphics[width=0.35\textwidth]{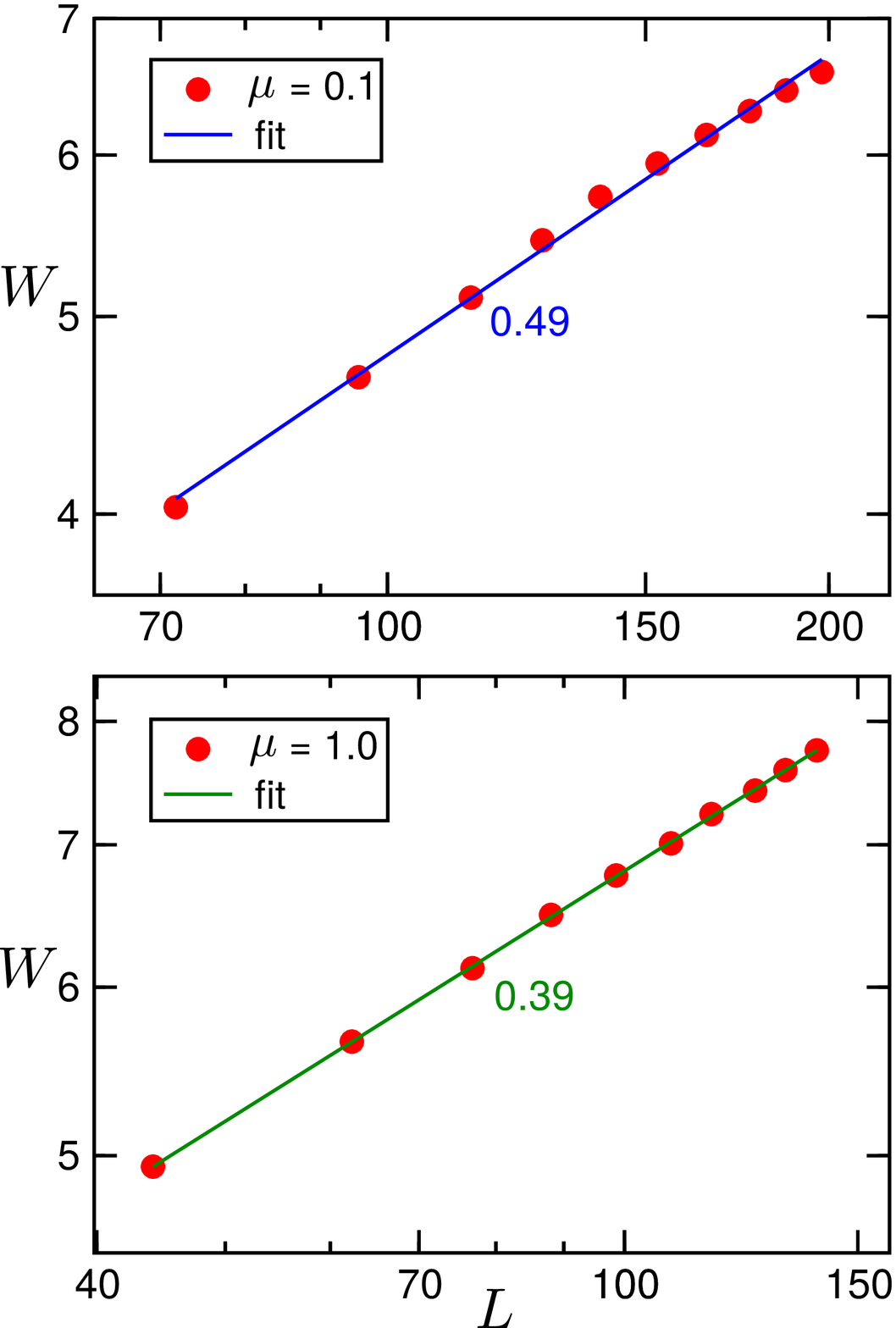}
\includegraphics[width=0.38\textwidth]{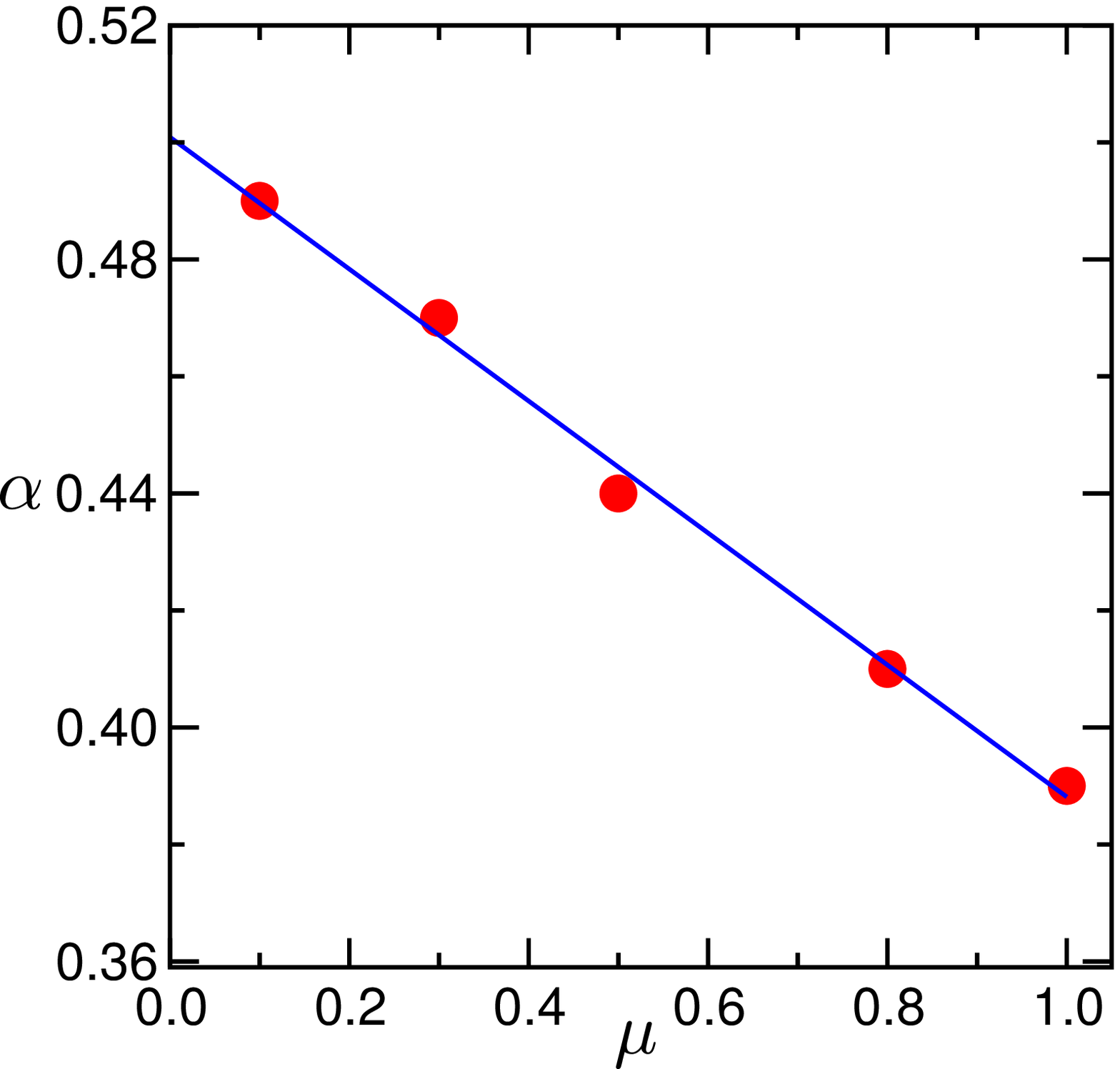}
\caption{Upper and middle panels: log-log plots of the width $W$ as a function 
of the length $L$
on the boundary of the sandpile. For different values of $\mu$ the scaling 
exponent $\alpha$ in
Eq.~(\ref{kpz}) changes. In the lower panel we show the values of $\alpha$ 
obtained for different
values of $\mu$. The exponent $\alpha$ appears to extrapolate to $\alpha=1/2$ 
for $\mu\to 0$.}
\label{WandL}
\end{figure}

Obviously in our model with the chosen parameters the scaling assumption can be 
broken. As said above, this can
stem from the existence of a number of unrelated typical length scales. In 
softer systems one could expect buckling
to introduce an important scale which we do not expect and do not observe in the 
present model. A second typical scale
can stem from the bifurcations of the force chains, cf. Fig.~\ref{forcechains}. 
This length had been studied in Ref.~\cite{01BCLO}. If the number density of 
bifurcation points is $n$ per unit area, then the probability $p$ that a 
particle lies at a bifurcation
point scales like $p\sim n\lambda^2$, and the bifurcation length $\xi_b\sim 
\lambda/p\sim 1/(n\lambda)$.
But in our opinion it is not the relevant length scale responsible for the 
breaking of scaling because it is
not growing with the system size. Thus it can renormalize the elastic properties 
of the medium but for large enough
piles we do not expect it to destroy the scaling behavior. On the other hand, 
the height fluctuations on the surface
of the pile do grow with the system size (albeit sub-extensively) and can be 
responsible for destroying the scaling
solutions also for large piles. The scaling exponents characterizing this scale 
were discussed in the literature,
cf. Ref~\cite{94BCPE} and references there in. We can however measure the height 
fluctuations over an edge of length $L$ with
growing sandpile sizes. We find that as a result
of the surface growth and re-construction the rms height fluctuations $W(L)$ 
scale like
\begin{equation}
W(L) \sim \lambda \left(\frac{L}{\lambda}\right)^\alpha \ .
\label{kpz}
\end{equation}
As can be seen from Fig.~\ref{WandL}, the value of the exponent $\alpha$ depends 
on the friction coefficient $\mu$, tending
to $\alpha\approx 0.5$ for small values of $\mu$. For larger values of $\mu$ the 
exponent $\alpha$ decreases, reaching a value
$\alpha\approx 0.39$ for $\mu=1$.

For a sandpile with $N$ disks and angle of repose $\phi$ the sandpile is 
approximately a triangle with
$z_{\rm max} = x_{\rm max} \tan\phi$. Elementary trigonometry then results in 
the estimate
\begin{equation}
z_{\rm max} \approx \lambda\sqrt{N\tan\phi}\ , \quad L\approx \lambda 
\sqrt{\frac{2N}{\sin 2\phi}} \ .
\label{LN}
\end{equation}
%%%%%%%%%%%%%%%%%%%%%%%%%%%%%%%%%%%%%%%%%%%%%%%%%%%%%%%%%%%%%%%%%

The ratio $\cal R$ of $W(L)$ to the height of the sandpile $z_{\rm max}$ will 
give us an estimate for the importance
of the surface fluctuation relative to the size of the sandpile. This ratio is
\begin{equation}
{\cal R} = \frac{W(L)}{z_{\rm max}} \approx \frac{2^{\alpha/2} 
N^{\frac{(\alpha-1)}{2}}}{(\sin2\phi)^{\alpha/2} \sqrt{\tan\phi}} \ .
\label{R}
\end{equation}
We now note that for small values of $\mu$ the angle of repose $\phi\to 0$ and 
for $\alpha\approx 1/2$,
\begin{equation}
{\cal R}\sim N^{-1/4} \phi^{-3/4} \ ,
\label{Rint}
\end{equation}
which becomes of the order of unity when the angle of repose reaches a critical 
value $\phi_c$ where
\begin{equation}
\phi_c\sim N^{-1/3} \ .
\label{predict}
\end{equation}
The meaning of this result is that for a fixed angle of repose there will be a 
value of $N$ below which scaling
is broken since ${\cal R}>1$. For small values of $\phi$ this can be a very 
large value of $N$.

The same conclusion can be reached from a different point of view. Fluctuations 
in the surface height will lead to
fluctuations in the angle of repose of standard deviation $\delta \phi$ where
\begin{equation}
\delta \phi\sim \frac{W}{L} \sim \left(\frac{L}{\lambda}\right)^{(\alpha-1)} 
\sim \left(\frac{2N}{\sin 2\phi}\right)^{-1/4} \ .
\label{finW}
\end{equation}
For small angle of repose the fluctuation in the angle of repose approach the 
magnitude of the angle of repose itself,
when the angle of repose $\phi_c$ is of the order $\phi_c\sim N^{-1/3}$ as had 
been previously estimated in
Eq.~(\ref{predict}). Note that for larger values of $\mu$ similar estimates can 
be made, but the angle
$\phi$ is no longer small, so nonlinear corrections are called for.
%%%%%%%%%%%%%%%%%%%%%%%%%%%%%%%%%%
\subsection{Consequences of breaking of scaling at the bottom}
\label{bottom}
%%%%%%%%%%%%%%%%%%%%%%%%%%%%%%%%%%%%%%%%%%%%%%%%%%%%%%%%%%%

Let us discuss the dip in pressure under the apex of the pile further. To do so 
we
need to focus on the function $\Psi(S,z=h)$. At the bottom when $S\to 1$ one 
always encounters an
edge of the pile with newly added grains that are highly susceptible to height 
fluctuations.
Therefore the form of $\Psi(S\to 1, z=h)$ is no longer given by the form 
Eq.~(\ref{psi1}). Since the
particles at the edge are mostly subject to gravity rather than the whole weight 
of a pile on their
shoulders, we expect
\begin{equation}
\Psi(1,z=h)=0
\label{psi2}
\end{equation}
to hold.
This dramatic change in the form of $\Psi(S,z=h)$ is plotted in 
Fig.~\ref{psilines}. Specifically, while $\Psi(S\rightarrow 0,z=h)\rightarrow 0$ 
 for both scaling and broken scaling solutions, the behavior for 
$\Psi(S\rightarrow 1,z=h)$ is very different for the scaling and broken scaling 
principal axis directions. This difference in the functional form of $\Psi$ is 
probably due to both the interaction of the sandpile with the surface on which 
it lies, as well as  fluctuations of the sandpile surface.

Note also that $\Psi(S\to 1, z=h)$, as can be also seen directly in the 
simulation results in
Figs.~\ref{psidir} and \ref{psilines},   that at some value of $S$, say $S_m$, 
inside the pile, this
function has a maximum which is of the order of the value given by the scaling 
solution Eq.~(\ref{psi1}).
The value of $S_m$ can be estimated as follows. Remembering that $S\equiv 
x\tan\phi/z$ we now use the
shift $\Delta x$ towards the maximum and estimate
\begin{equation}
S_m =1 - \frac{\Delta x\tan\phi}{h} \sim 1-\frac{W}{L} \ ,
\end{equation}
where we have used $\Delta x= W\cos\phi$ and $h=L\sin\phi$.
Using now Eq.~(\ref{finW}) we end up with the estimate
\begin{equation}
S_m =1-\left(\frac{2N}{\sin 2\phi}\right)^{-1/4} \ .
\label{finsm}
\end{equation}
In our simulations with $N$ of the order of 1000 $S_m\approx 0.8$ in very good 
agreement with the
numerical results.

\begin{figure}
\includegraphics[width=0.38\textwidth]{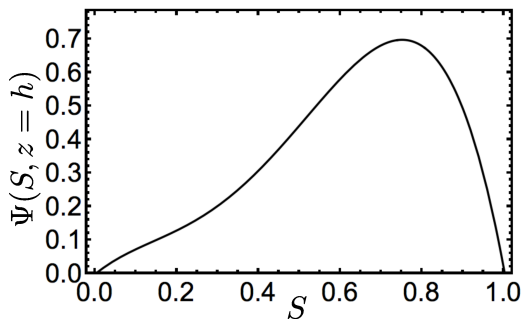}
\caption{A model function $\Psi(S,z=h)$ which mimics the behavior of the 
function found in the numerics
near the bottom of the pile, as shown in Fig.~\ref{psilines}.}
\label{half}
\end{figure}

\begin{figure}
\includegraphics[width=0.38\textwidth]{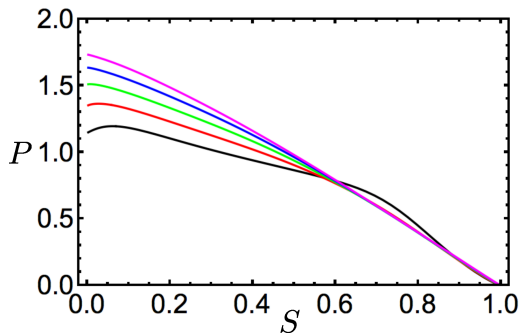}
\caption{The pressure at the bottom of the pile obtained by integrating the 
continuum equations
starting from the edge. The different curves are for for the same values of 
$\phi$
as in Fig.~\ref{numsol}. For larger angles of
repose a dip in the pressure is developing naturally. }
\label{dip}
\end{figure}

Accordingly we present in Fig.~\ref{half} the positive branch of an 
anti-symmetric function $\Psi(S,z=h)$ that begins at zero and ends up at
zero with a maximum at the expected value of $S_m$.
%%%%%%%%%%%%%%%%%%%%%%%%%%%%%%%%%%%%%%%%%%%%%%%%%%%%%%%%%%%%%%%%%
%%%%%%%%%%%%%%%%%%%%%%%%%%%%%%%%%%%%%%%%%%%%%%%%%%%%
This function is simply a fourth order polynomial
with coefficients chosen to respect all the wanted properties.
The reader should compare this function with ``bottom" function shown in the 
lower
panel of Fig.~\ref{psilines}. Plugging this function as the constitutive 
relation into Eqs.~(\ref{final}), (\ref{constit}),
we can integrate the equations starting from the outer edge.  Solving for the 
pressure $P\equiv s_{xx}+s_{yy}$ as
a function of the angle of repose we get the curves shown in Fig.~\ref{dip}.

%%%%%%%%%%%%%%%%%%%%%%%%%%%%%%%%%%%%%%%

 Note that we do not need to
match two solutions starting from the edge and from the center. The conclusion 
is that when
the angle of repose is small, the pressure is expected to maximize under the 
apex of the pile. For increasing angles
of repose (equivalently increasing $\mu$) there is an increasing tendency for 
the pressure to dip under the apex
with a maximum away from the center. The results indicate that the reason for 
the dip within the classical
continuum theory is actually the breaking of the scaling assumption by the 
existence of a bottom boundary to the
sand-pile. Note that this breaking of scaling is different from the one 
discussed in the previous subsection
where surface fluctuations introduced a typical length-scale.

%%%%%%%%%%%%%%%%%%%%%%%%%%%%%%%%%%%%%%%%%%%%%%%%%
\section{Summary and discussion}
\label{summary}
In this paper we have examined sandpiles that were grown using a new 
quasi-static model
which allows a detailed computation of the characteristics of the sandpile, 
including its
angle of repose, the stress field, the orientation of the principal axis of the 
stress etc.,
all as a function of the coefficient of friction and the system size. We can 
compare these results
with the predictions of a theory based on continuum equilibrium mechanics. As 
explained, the
equations of equilibrium mechanics are under-determined and their solution 
requires and input of
constitute relations. In addition, seeking scaling solutions leads to a very 
elegant formalism,
which appears however to be challenged by the numerical simulations. Note that 
even when scaling is obeyed,
the fact that $\Psi$ is a function of $S$ implies memory in the growth of the 
sandpile. Further, our sandpiles have
stress fields which do not agree with the scaling assumption in all detail, 
resulting in the function
$\Psi$ depending on both $S$ and $z$ and not only on $S$. Fortunately, for 
larger values of the friction
coefficient and for larger system sizes the breaking of scaling is weak in the 
bulk, allowing an approximate analytic theory
which agrees well with the observations. On the other hand, scaling is strongly 
broken even for large $N$ and for
large $\mu$ at the bottom of the pile near the floor. Interestingly enough, when 
we input the data for $\Psi$ found in the numerics into the analytic theory, we 
find the often observed dip in the pressure at the center of the pile, {\em 
without
needing to match two piece-wise linear solutions as was necessary in previous 
publications}, \cite{96WCCB,97WCC}. We propose that our model provides evidence 
that the dip in the pressure results from the broken scaling solution which 
presumably is
generic due to the special interactions of the particles with the substrate 
support.

The mechanism proposed here for the appearance of a typical scale that breaks 
the scaling solutions is not the only one possible. Another interesting 
proposition was offered in Ref.~\cite{01BCLO} where the length-scale associated 
with the bifurcations of the force chains leads to a convective-diffusive 
equation for the stress field. Solution of this
equation predict a dip in the pressure at the bottom of the pile. It is possible 
that there exist other mechanisms. In fact, one should stress that the 
particular properties of the numerical model employed
in this paper may very well affect the solution. Changing the dissipation 
mechanism associated with frictional slips, replacing the quasi-static growth of 
the sandpile with continuous additions of grains before mechanical equilibrium 
is restored, and other such details, may result in very different profiles of 
$\Psi(S,z)$. Thus one
cannot judge agreement or disagreement with other models \cite{14ZY,05GG}of the 
sandpile without comparing the profiles of $\Psi(S,z)$. In the present case 
discussed above the appearance of non-scaling solutions is strongly supported by 
the numerics, and the agreement with their shape as discussed in 
Sect.~\ref{bottom} gives credence to the mechanism proposed here. It is not 
impossible that details of the grains shapes, their interactions between 
themselves and also with the floor
may require different approaches to explain the observed characteristics of the 
sand pile. What should
remain invariant is the approach proposed here to expand the solutions near the 
center of the pile in
accordance with the measured profile of $\Psi$ to find what the theory 
predicts. 

\acknowledgments
This work had been supported by the Minerva foundation with funding from the 
Federal German Ministry for Education and Research.
PKJ is grateful to the VATAT fellowship from the Council of Higher Education, 
Israel.

\end{document}